\documentclass[a4paper]{revtex4}
\pdfoutput=1
\usepackage{amssymb}
\usepackage{amsmath,upgreek,soul}
\usepackage{multirow}
\usepackage{graphicx}
\usepackage{subfig}
\usepackage{adjustbox}
\usepackage[amssymb,cdot]{SIunits}

\def\znu2b{$0\nu\beta\beta$}
\usepackage[normalem]{ulem}

\begin{document}

%%% \markboth{Cheng et al.}{CJPL Early Science}

\title{The China Jinping Underground
Laboratory and Its Early Science
$^($\footnote{Published in
Annual Review of Nuclear and Particle Science {\bf 67},
231-251 (2017).}$^,$\footnote{DOI: 10.1146/annurev-nucl-102115-044842}$^)$
}
\author{Jian-Ping Cheng}
\author{Ke-Jun Kang}
\author{Jian-Min Li}
\author{Jin Li}
\author{Yuan-Jing Li}
\author{Qian Yue}
\altaffiliation[Corresponding Author: ]{ yueq@mail.tsinghua.edu.cn }
\author{Zhi Zeng}
\altaffiliation[Corresponding Author: ]{ zengzhi@tsinghua.edu.cn }
\affiliation{Key Laboratory of Particle and Radiation Imaging
  (Ministry of Education) and Department of Engineering Physics,
  Tsinghua University, Beijing 100084 }
\author{Yun-Hua Chen}
\author{Shi-Yong Wu}
\affiliation{Yalong River Hydropower Development Company, Chengdu
  610051}
\author{Xiang-Dong Ji}
\affiliation{INPAC and Department of Physics and Astronomy, Shanghai
  Laboratory for Particle Physics and Cosmology and Shanghai Jiao Tong
  University, Shanghai 200240}
\author{Henry T. Wong}
\altaffiliation[Corresponding Author: ]{ htwong@phys.sinica.edu.tw }
\affiliation{Institute of Physics, Academia Sinica, Taipei 11529 }

\begin{abstract}

  The China Jinping Underground Laboratory, inaugurated in 2010, is an
  underground research facility with the deepest rock overburden and largest
  space by volume in the world. The first-generation science programs include
  dark matter searches conducted by the CDEX and PandaX experiments. These
  activities are complemented by measurements of ambient radioactivity and
  installation of low-background counting systems. Phase II of the facility is
  being constructed, and its potential research projects are being
  formulated. In this review, we discuss the history, key features, results, and
  status of this facility and its experimental programs, as well as their future
  evolution and plans.\\
\textbf{keywords:} underground laboratory, low-background
  instrumentation, dark matter searches, neutrino physics
\end{abstract}

%%\keywords{
%%  underground laboratory, low-background instrumentation, dark
%%  matter searches, neutrino physics
%%}

\maketitle

\tableofcontents{}

\section{INTRODUCTION}
\label{sect::intro}

An established approach to the search for new physics involves
experimental studies of rare-occurrence events. This typically
requires low-background experiments with large target mass and
elaborate shielding to suppress ambient radioactivity due to photons
and neutrons. Direct and induced effects due to cosmic rays, however,
cannot be shielded. Accordingly, such experiments should be performed
in laboratories located underground, where a kilometer-scale rock
overburden can effectively attenuate the cosmic-ray fluxes and their
induced background.

Many underground facilities have been constructed around the world and
house a range of diversified science
programs~\cite{Lesko:2015aip}. Proton decays are generic predictions
of Grand Unified Theories~\cite[pp.270--78]{RPP2014} and attracted a
great deal of interest in underground experiments and their techniques
in the 1970s. This early phase was followed by a series of important
results on neutrino oscillations and
mixings~\cite[pp.235--58]{RPP2014} produced by underground
experiments, which provided the first glimpse of physics beyond the
Standard Model. Ongoing and future generations of long-baseline
neutrino beam experiments, which also involve massive underground
detectors, will advance the study of the neutrino mixing matrix. Also
under way are experimental studies of neutrinoless double-$\beta$
decay \cite[pp.698--703]{RPP2014}, which aim to demonstrate that
neutrinos are Majorana particles. Beyond neutrino physics, underground
laboratories are also home to an array of experiments performing
direct searches for dark matter ~\cite[pp.353--60]{RPP2014} in the
context of \MakeLowercase{Weakly Interacting Massive Particles}
(WIMPs, denoted by $\chi$) and axions.

A new underground facility inaugurated in December 2010---the China
Jinping Underground Laboratory
(CJPL)~\cite{CJPL:2010,CJPL:2015}---will add high-quality laboratory
space to this worldwide program. In this review, we present an
overview of the first-generation efforts (Phase I) at CJPL. We discuss
the first science programs, particularly CDEX (\uline{C}hina
\uline{D}ark Matter \uline{Ex}periment;
~\cite{CDEX:2013,Yue:2016jpcs}) and PandaX~(Particle {\underline{and}}
{\underline{A}}strophysical {\underline{X}}enon;
\cite{PANDAX:2014ax}), and present their results. We then describe the
associated ambient background measurements and low-counting-rate
facilities at CJPL~\cite{Zeng:2014ar}. We conclude with a discussion
of the status and plans of Phase II~\cite{CJPL:2015}, which is
currently under construction.

\section{CHINA JINPING UNDERGROUND LABORATORY: PHASE I}
\label{sect::cjpl1}

Since the 2000s, interest in constructing hydro-power facilities in the Yalong
River area of Sichuan Province, China, has intensified \cite{Wu:2010sn}. In
2008, two 17.5-km-long traffic tunnels under Jinping Mountain were completed to
facilitate the transport of construction materials.  The physics community in
China immediately recognized the opportunities and potential of this
development. By 2009, Tsinghua University and Yalong River Hydropower
Development Company reached an agreement to jointly develop an underground
laboratory facility---CJPL.

\begin{figure}[ht]
  \includegraphics[width=0.9\textwidth,scale=0.5]{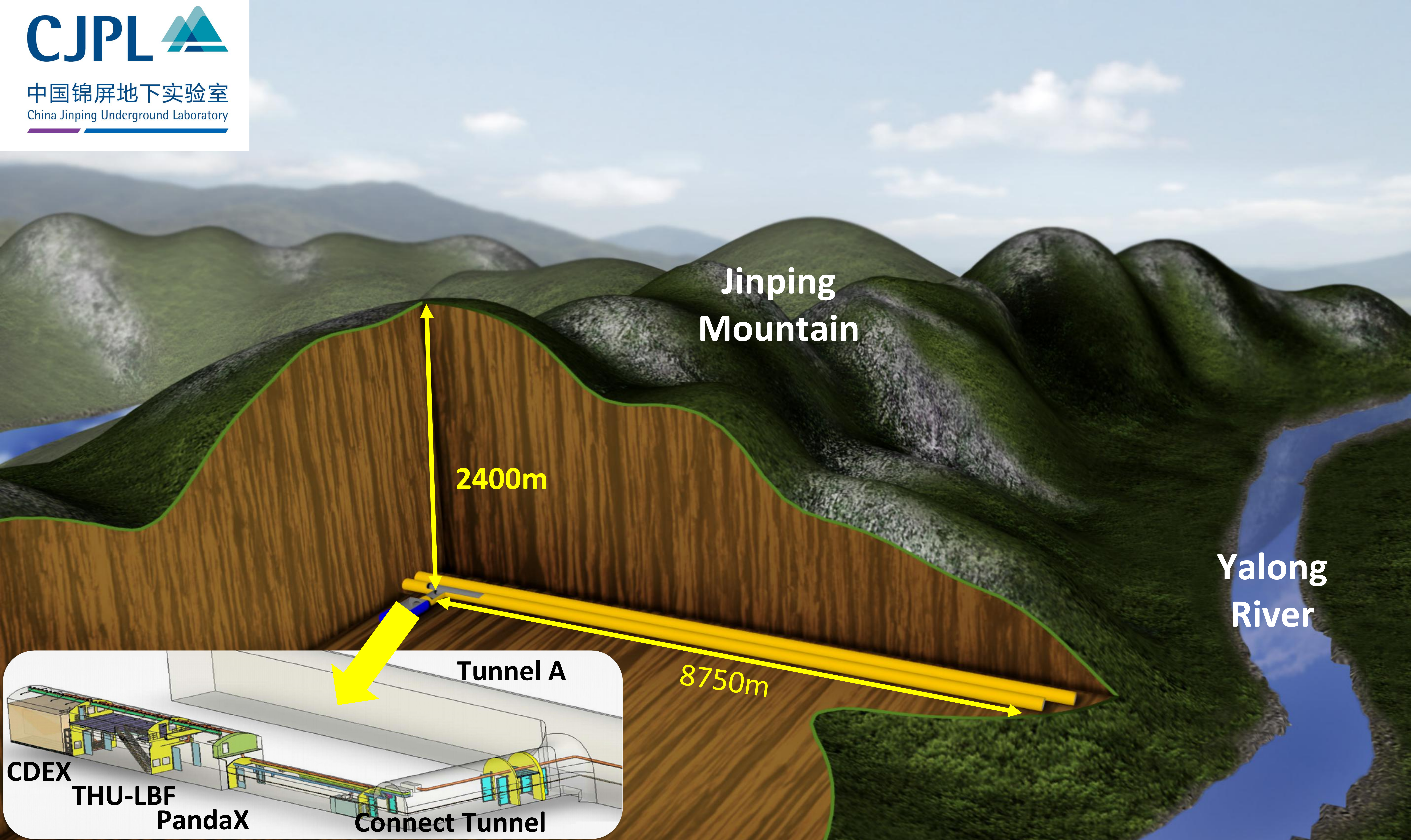}
  \caption{Schematic layout of CJPL-I, showing its geographical
    features and the location of the CDEX and PandaX experiments, as
    well as the low-background screening facilities (THU-LBF). These
    facilities are located in the main experiment hall, with
    dimensions of 6.5 m (width) $\times$ 6.5 m (height) $\times$ 40 m
    (length).}
  \label{fig::cjpl1_schema}
\end{figure}

Construction of Phase I of CJPL (CJPL-I)\ started in December 2009 near the
center of one of the Jinping traffic tunnels, where the vertical rock overburden
is maximal. The inauguration of the facility took place in December
2010. Table~\ref{tab::cjpl_pars} summarizes the key features of CJPL-I and
presents a comparison between this facility and the similar Gran
Sasso~\cite{GRANSASSO:2012ep} and SNOLab~\cite{SNOLAB:2012ep} underground
facilities. Figure~\ref{fig::cjpl1_schema} presents a schematic diagram of
CJPL-I.

%%  ======  Table I =============

\begin{table}[ht]
  \caption{Key features of China Jinping Underground Laboratory, Phase
    I\ (CJPL-I), and comparison with benchmark underground facilities}
  \label{tab::cjpl_pars}
\begin{center}
  \begin{tabular}{llllllll}
    \hline
    Facility & Location & Host & Access & Volume & Rock burden/ & Muon flux & Start \\
             & & structure & & (m$^3$) & depth (m) & (m$^{-2}$ day$^{-1}$) & year  \\
    \hline
    SNOLab \cite{SNOLAB:2012ep} & Canada & Mine & Shaft & 30,000 & 2,000 & 0.27 & 2004 \\
    Gran Sasso \cite{GRANSASSO:2012ep} & Italy & Traffic tunnel
                               & Drive-in  & 180,000 & 1,400 & 26 & 1987 \\
    \hline
    \multicolumn{8}{l}{CJPL \cite{CJPL:2010,CJPL:2015} } \\
    CJPL-I  & {China} & {Traffic tunnel} & Drive-in & 4,000   & 2,400 & 0.17 & 2010  \\
    CJPL-II &   & & Drive-in & 300,000 & 2,400 & & 2017 \\
    \hline
  \end{tabular}
\end{center}
\end{table}

%%  ===================

CJPL-I was built under Jinping Mountain with 2,400 m of rock
overburden. It has drive-in access via a two-lane road tunnel with
enough headroom for construction trucks. (The distances to the east
and west ends of the tunnel entrance are 8.7 km and 8.8 km,
respectively.) The main hall of CJPL-I, where the experiments were
installed, has dimensions of 6.5 m (width) $\times$ 6.5 m (height)
$\times$ 40 m (length), and accordingly the floor area is 260 m$^2$.

\begin{figure}[ht]
  \includegraphics[width=0.9\textwidth]{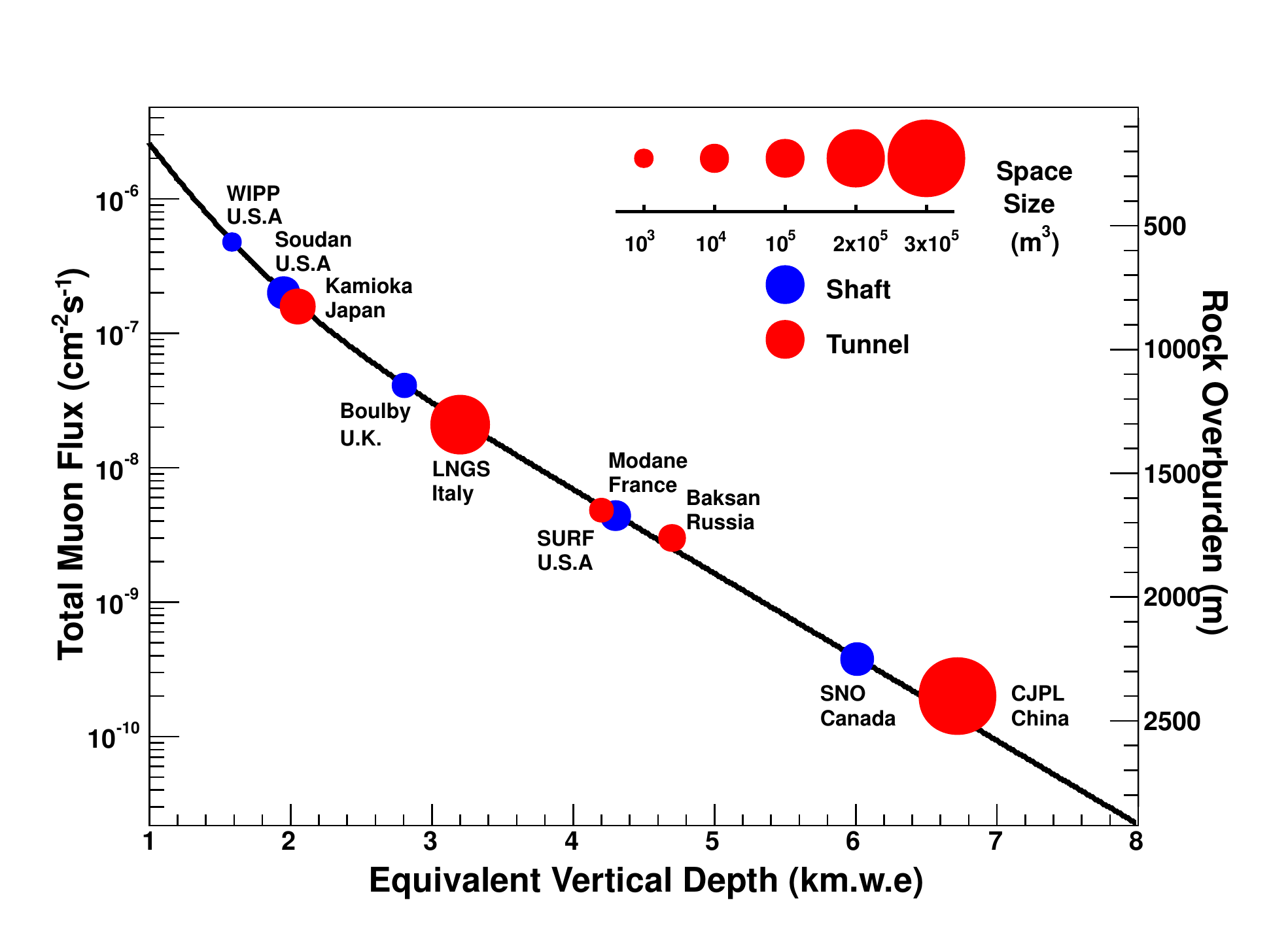}
  \caption{The measured residual muon fluxes in key underground
    facilities, which are consistent with predicted values
    (\textit{black line}). The sizes of the circles correspond to
    laboratory space by volume; red or blue denotes access by road
    tunnels or shafts, respectively.}
  \label{fig::cjpl_mflux}
\end{figure}

The facility is the deepest operating underground laboratory in the
world. The measured fluxes of residual cosmic-ray
muons~\cite{Wu:2013cp} at CJPL-I, and a comparison with various active
underground laboratories, are displayed in
Figure~\ref{fig::cjpl_mflux}. We note for completeness that early
cosmic-ray measurements had been performed at the Kolar Gold Fields in
India, at locations with 2,760 m of rock
overburden~\cite{Narasimhan:2004in}.

International access to CJPL is by regular flights from various hub
cities in China to Xichang Domestic Airport. The laboratory is located
90 km from the airport and is reachable by both highways and two-lane
paved private roads shared with Yalong River Hydropower Development
Company; the driving time is approximately 2~hours.  A\ guest house,
dormitory accommodation, canteens, and office space are available near
the west entrance of the tunnel.

The bedrock surrounding CJPL is made of marble, with relatively low
radioactivity from the contamination of $^{232}$Th and $^{238}$U
isotopes (measurements of these isotopes and other background
characterizations are discussed in Section~\ref{sect::lobkgf}). The
laboratory space is shielded from the bare rock by a 0.5-m-thick layer
of concrete layer. Ventilation is provided by a 10-km-long,
55-cm-diameter pipeline that brings in fresh air from the west
entrance of the tunnel at an air-exchange rate of 4,500 m$^3$
h$^{-1}$, allowing the radon level to be kept below $20$ Bq m$^{-3}$.

The laboratory is connected by internet cables at a bandwidth of 10
GHz. The storage capacity of the computer servers, located in an
exterior office, is up to 300~TB, with a transfer capacity of 10~GHz
from the experiments. The laboratory is equipped with surveillance CCD
cameras and audiovisual alarm systems, and ambient conditions
including temperature, pressure, humidity, and air composition are
continually recorded. All recording devices are accessible from and
can be monitored at remote sites.

The CJPL-I laboratory space is shared among three scientific programs:
the CDEX~\cite{CDEX:2013} and PandaX~\cite{PANDAX:2014ax} dark matter
experiments and the Tsinghua University low-background facilities
(THU-LBF) for low-radiopurity measurements~\cite{Zeng:2014ar}. The
status and early results of these programs are discussed in the
following sections.

\section{CDEX }
\label{sect::cdex}

Approximately one-quarter of the energy density of the Universe can be
attributed to cold dark matter~\cite[p. 353]{RPP2014}, whose nature
and properties are unknown. WIMPs are leading candidates. They
interact with matter predominantly via elastic scattering with nuclei:
$\chi + N \rightarrow \chi + N$. The unique merits of CJPL make it an
ideal location in which to perform low-background experiments on dark
matter searches.

Germanium detectors sensitive to sub-keV recoil energy are a possible
means of probing so-called light WIMPs with mass 1 GeV
$< {m_{\chi}} <$ 10 GeV \cite{Yue:2004he,Wong:2006jp}. This
observation inspired detailed studies of \textit{p}-type point-contact
germanium detectors (pPCGe) with modular mass on the kilogram
scale~\cite{Luke:1989it,Barbeau:2007jc,Soma:2016nima}, which were
followed by various experimental efforts
~\cite{CoGeNT:2011pr,CoGeNT:2013pr,Lin:2009pr,Li:2013pr,Giovanetti:2015pp}.
The scientific goal of CDEX \cite{CDEX:2013}, one of the two founding
experimental programs at CJPL, is to pursue studies of light WIMPs
with pPCGe.

\subsection{First-Generation CDEX Experiments}

\begin{figure}[ht]
  \includegraphics[width=0.9\textwidth]{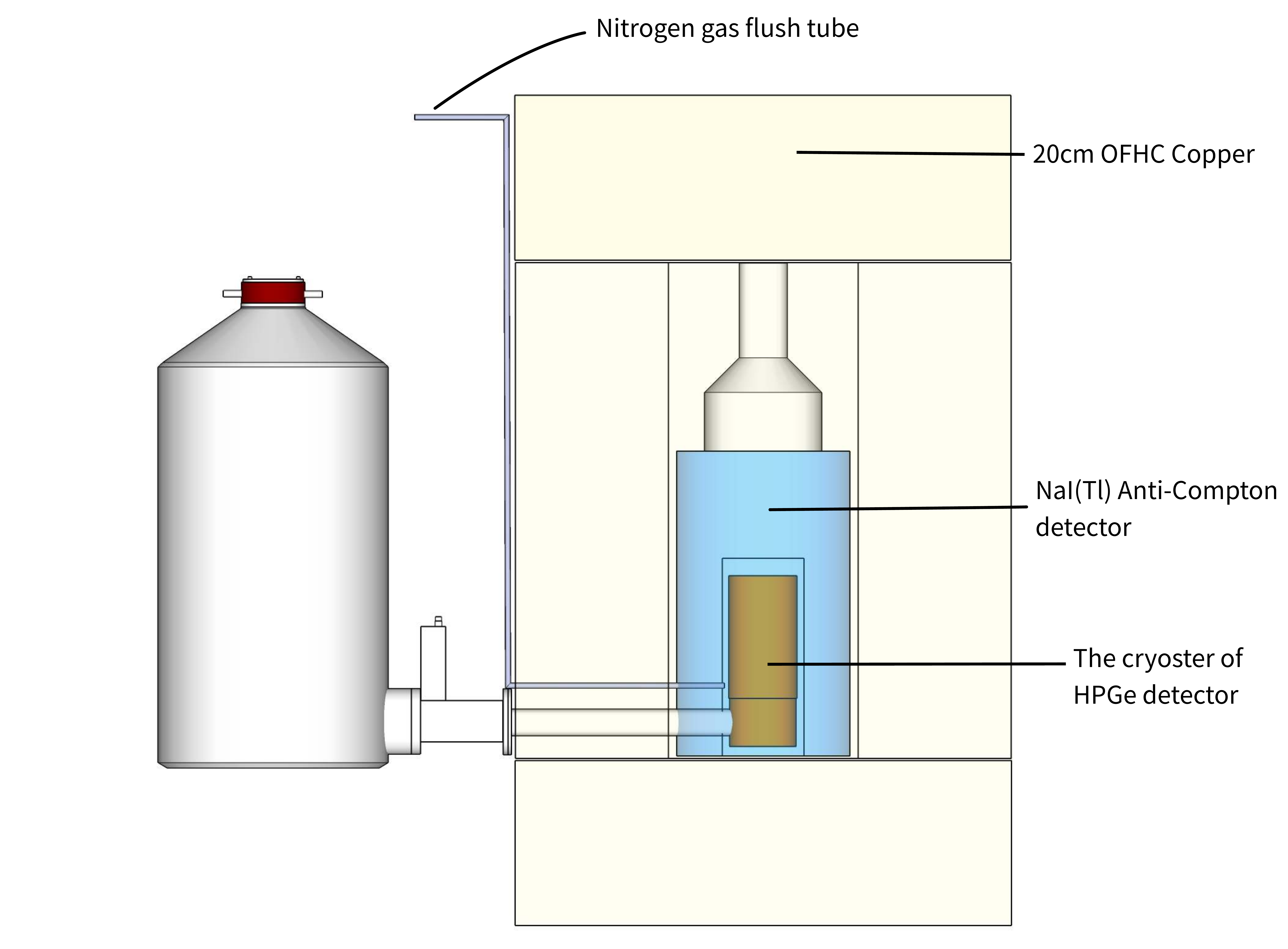}
  \caption{Schematic diagram of the baseline design of the CDEX-0 and
    CDEX-1 experiments, which use a \textit{p}-type point-contact
    germanium detector enclosed by anti-Compton NaI(Tl) crystal
    scintillator and passive shielding.  This setup is further
    shielded by an external structure (not shown) made of lead and
    copper layers. The entire shielding and detector system is
    installed inside a room with 1-m-thick polyethylene
    walls. Abbreviations: HPGe, high-purity germanium detector; OFHC,
    oxygen-free high-conductivity.}
  \label{fig::cdex1_setup}
\end{figure}

The first-generation CDEX experiments adopted a baseline
design~\cite{Lin:2009pr,Li:2013pr} of single-element 1-kg mass scale
pPCGe enclosed in NaI(Tl) crystal scintillator as an anti-Compton (AC)
detector (Figure~\ref{fig::cdex1_setup}). These active detectors are
surrounded by passive shielding made of oxygen-free high-conductivity
(OFHC) copper, boron-loaded polyethylene, and lead. The setup is
further shielded by additional OFHC copper and lead layers. The entire
shielding and detector system is installed inside a room with interior
dimensions of 4 m $\times$ 8 m $\times$ 4 m (height) and walls
constructed with 1-m-thick polyethylene.

\begin{figure}[ht]
  \begin{center}
    \hspace*{1.5cm}
    \subfloat[]{\includegraphics[width=0.6\linewidth]{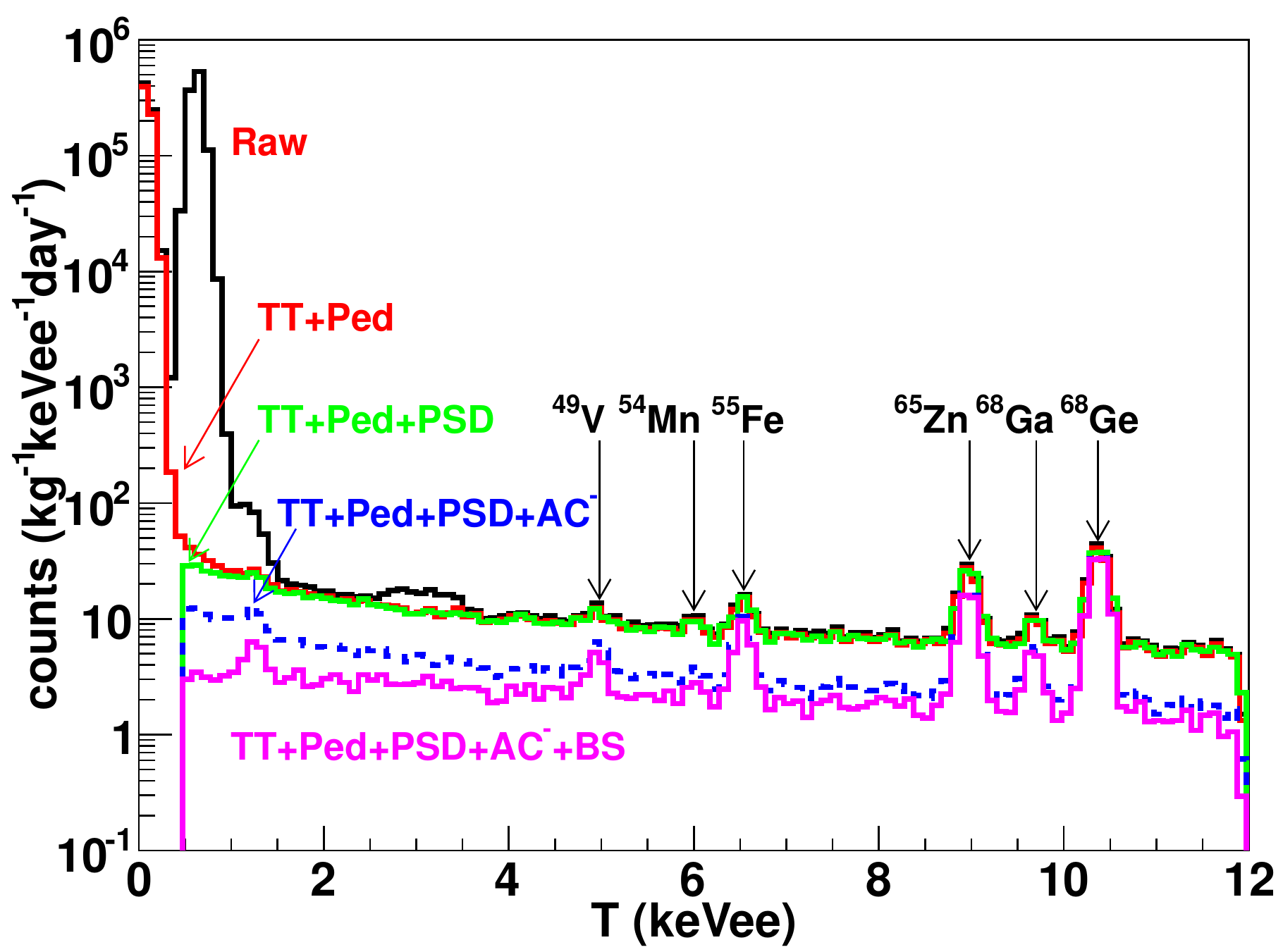}}
    \newline
    \subfloat[]{\includegraphics[width=0.42\linewidth]{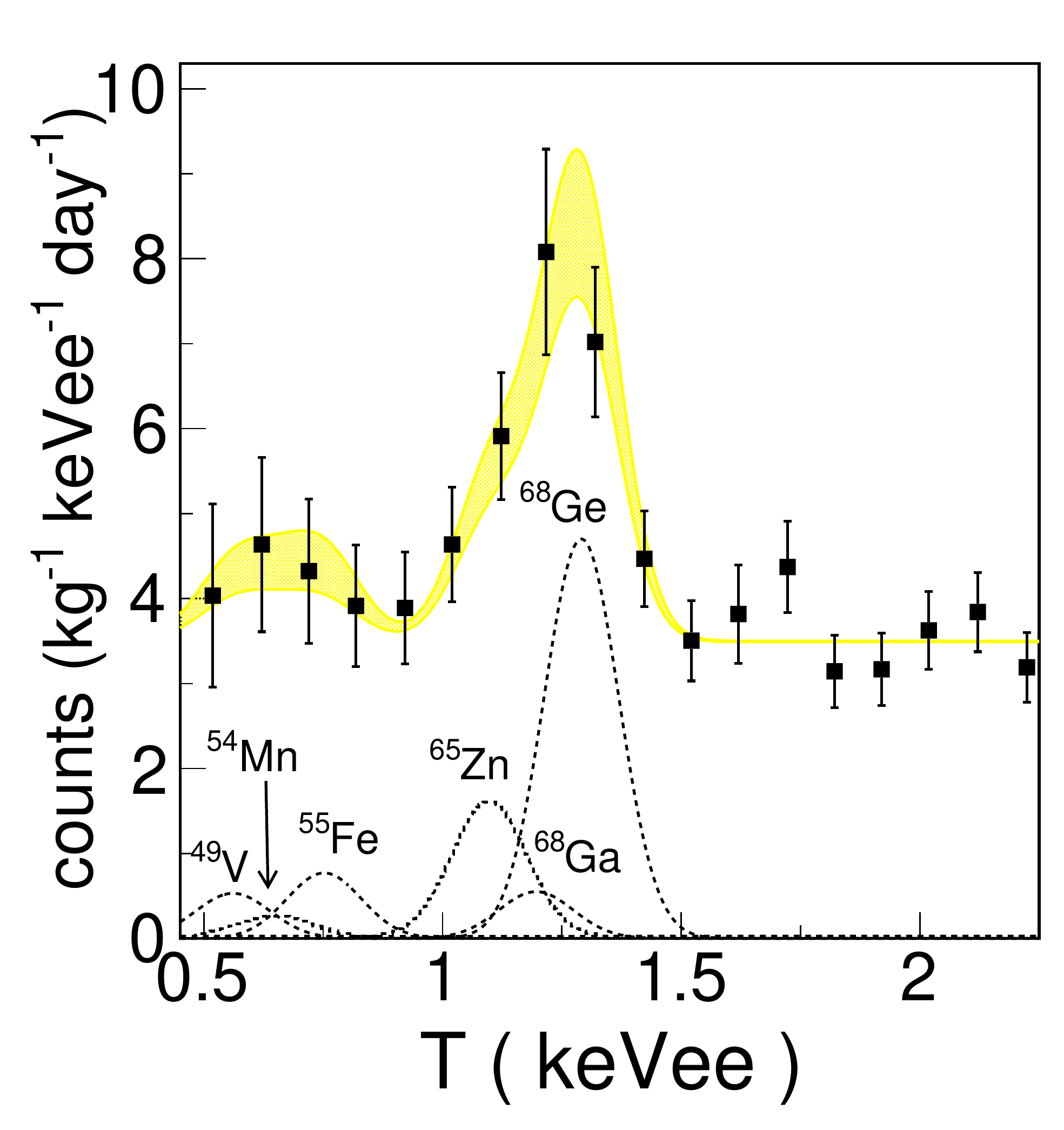}}
    \hspace*{1cm}
    \subfloat[]{\includegraphics[width=0.42\linewidth]{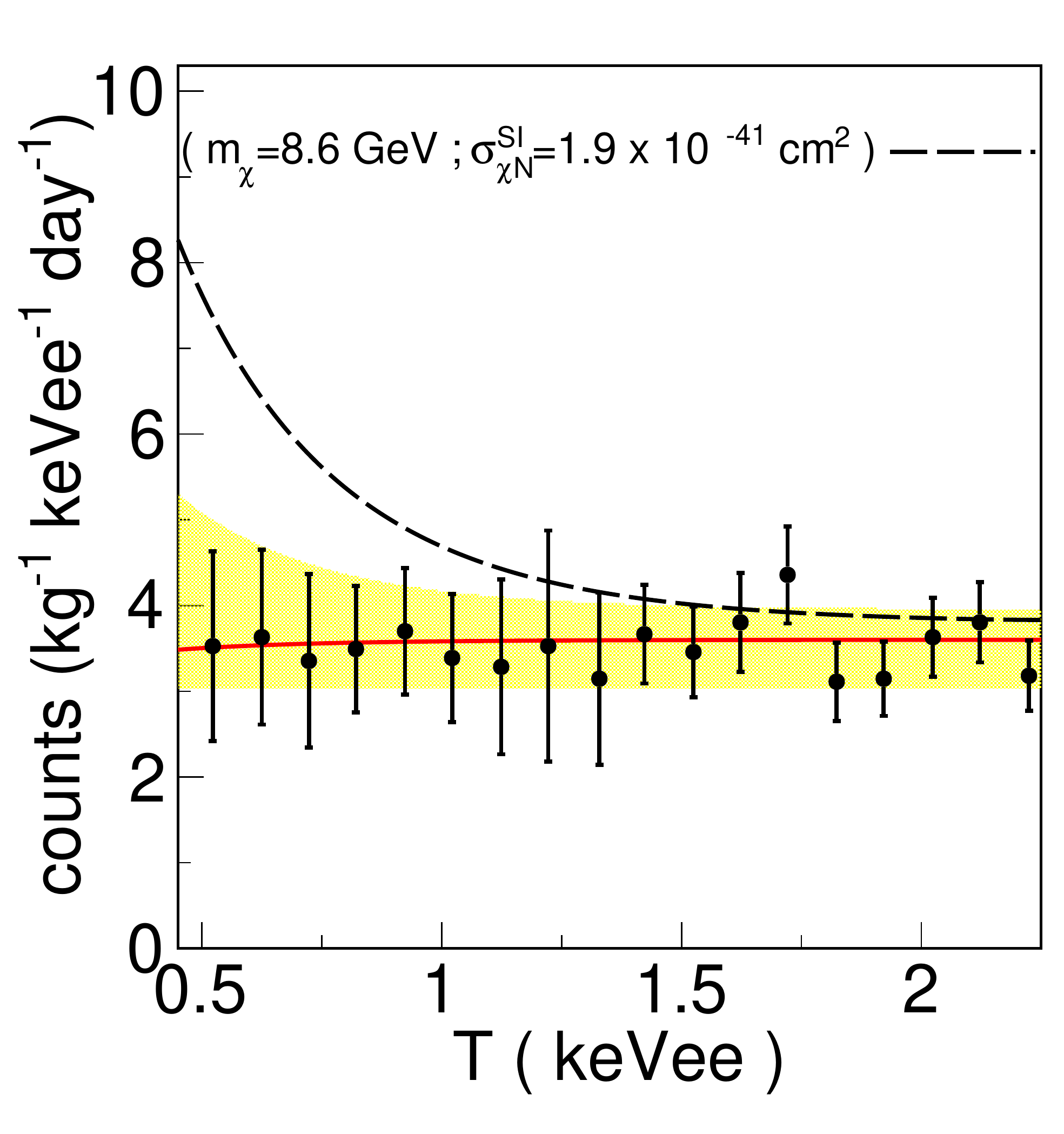}}
  \end{center}
  \caption{
Background spectra of the CDEX-1 experiment with
    335.6~kg$\cdot$day of exposure~\cite{Zhao:2016pr}.
    (\textit{a}) Spectra at various stages of candidate event
    selection: basic cuts (TT+Ped+PSD), anti-Compton (AC), and bulk
    (BS) events. The peaks correspond to internal X-ray emission from
    cosmogenically activated long-lived isotopes.
    (\textit{b}) Low-energy candidate events can be accounted for with
    known background channels: L-shell X-rays from internal cosmogenic
    radioactivity and a flat background due to ambient
    high-energy $\gamma$-rays.
    (\textit{c}) The residual spectrum superimposed with a
    hypothetical $\chi N$ recoil spectrum in an excluded parameter
    space.
}
  \label{fig::cdex1_spec}
\end{figure}

The pilot CDEX-0 measurement was based on a 20-g prototype germanium
detector at a 177-eV$_{ee}$ threshold with an exposure of
0.784~kg$\cdot$day~\cite{Liu:2014pr}. The CDEX-1 experiment adopted a
pPCGe detector with a mass of 1 kg. The first results [without the
NaI(Tl) AC detector] were based on an analysis threshold of 400
eV$_{ee}$ with an exposure of
14.6~kg$\cdot$day~\cite{Kang:2013cp,Zhao:2013pr}.  Subsequent data
taking incorporated the AC detector. After suppression of the
anomalous surface background events and measurement of their signal
efficiencies and background leakage factors with calibration
data~\cite{Li:2014ap,Yue:2014pr}, all residual events were accounted
for by known background models.  The sensitivities were further
improved with a longer exposure of 335.6~kg$\cdot$day, and new
constraints on spin-independent and spin-dependent $\chi N$ couplings
were derived~\cite{Zhao:2016pr}.  The results
(Figure~\ref{fig::cdex1_spec}) represent the most sensitive
measurements made with the point-contact high-purity germanium (HPGe)
detector.

\begin{figure}[ht]
  \begin{center}
    \subfloat[]{\includegraphics[width=0.45\linewidth]{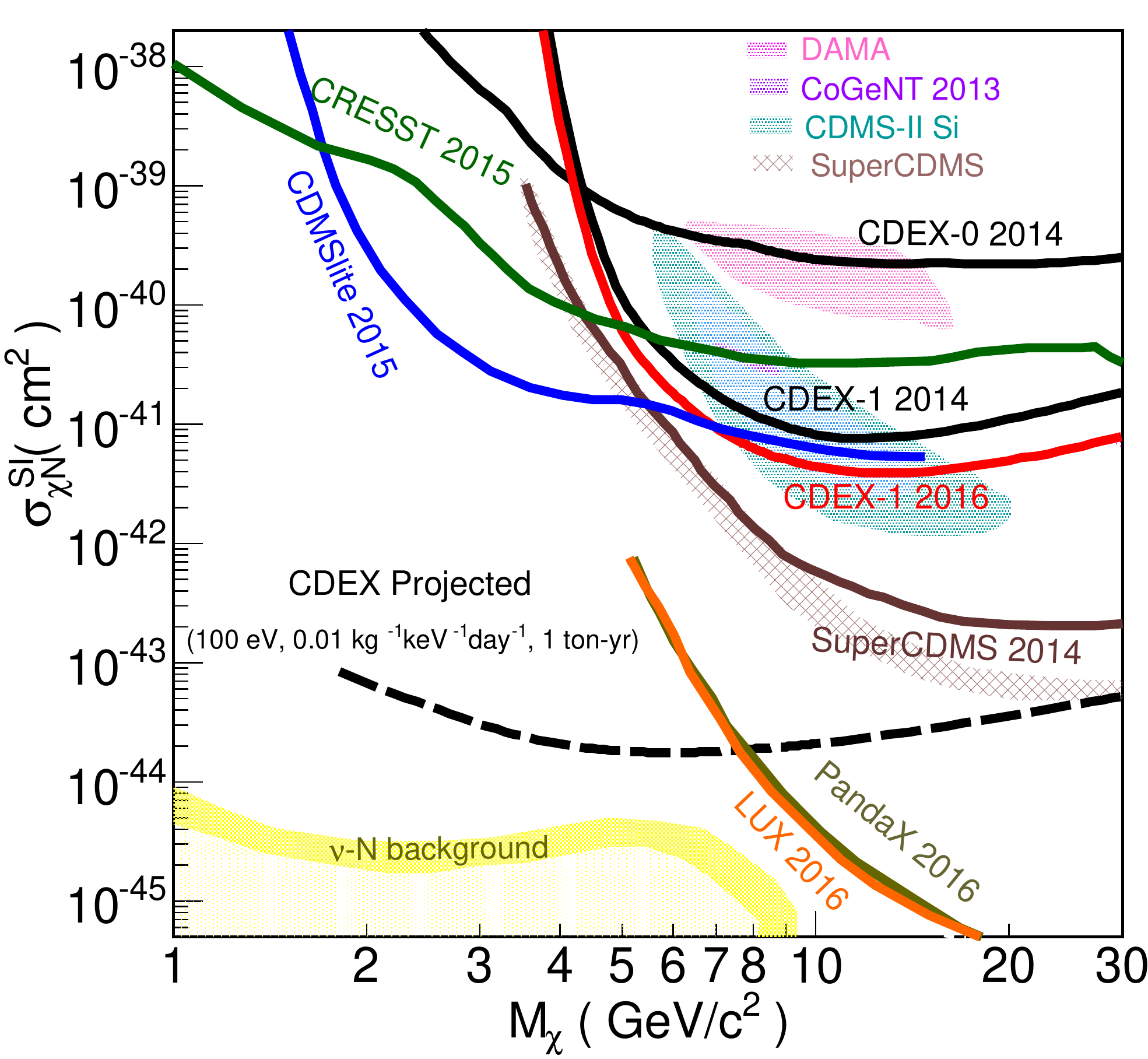}}
    \subfloat[]{\includegraphics[width=0.45\linewidth]{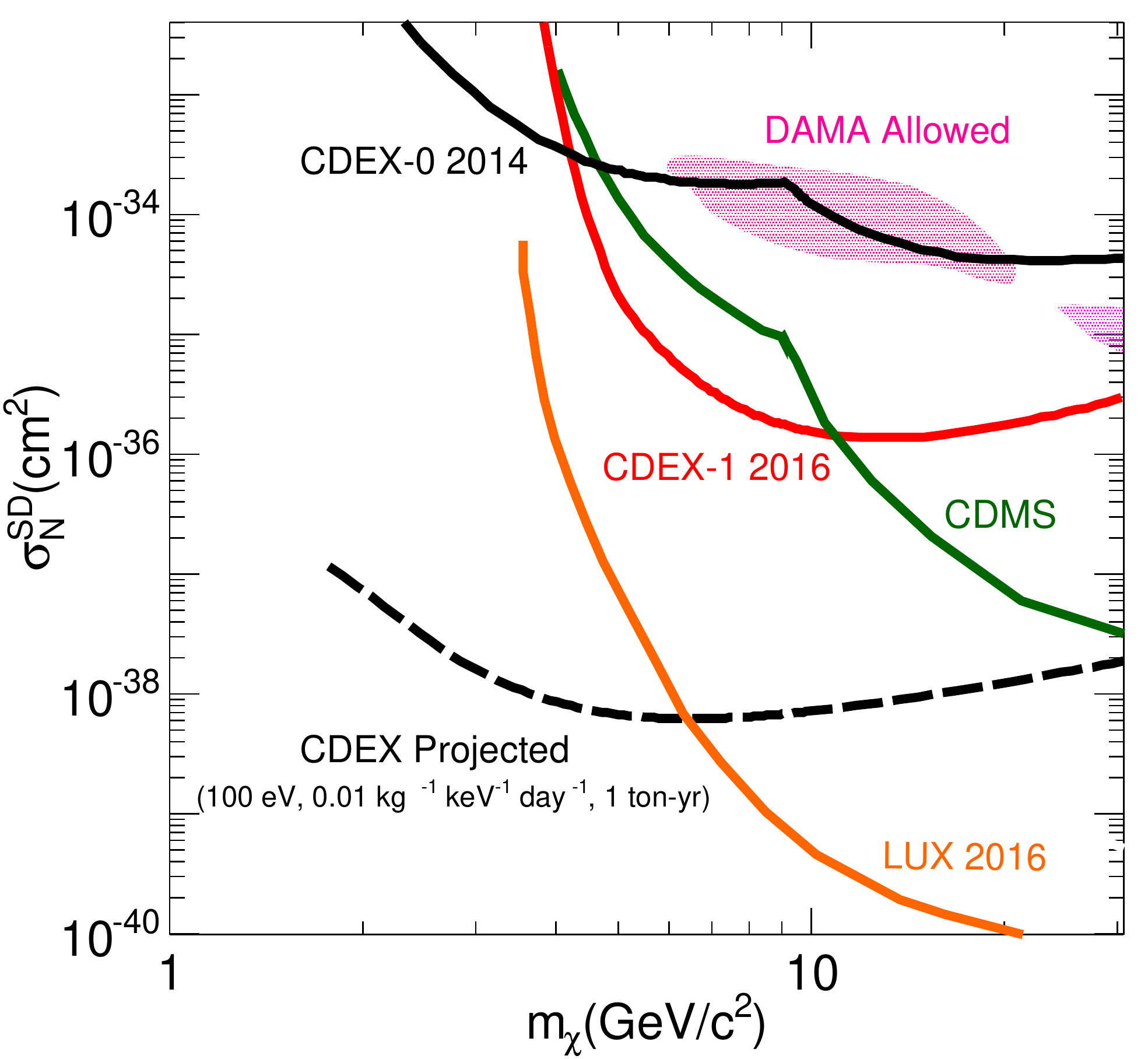}}
    \newline
    \subfloat[]{\includegraphics[width=0.45\linewidth]{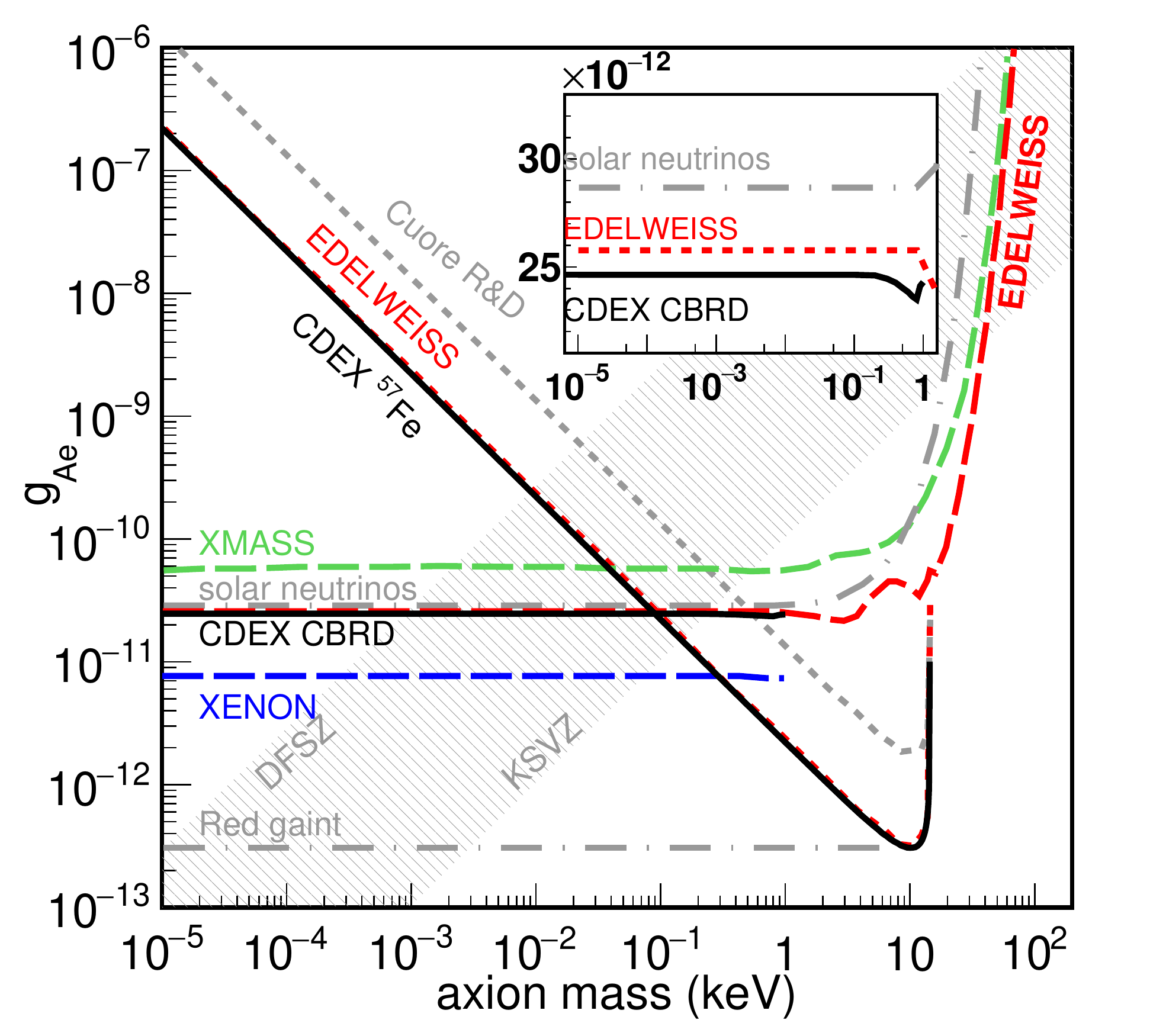}}
    \subfloat[]{\includegraphics[width=0.45\linewidth]{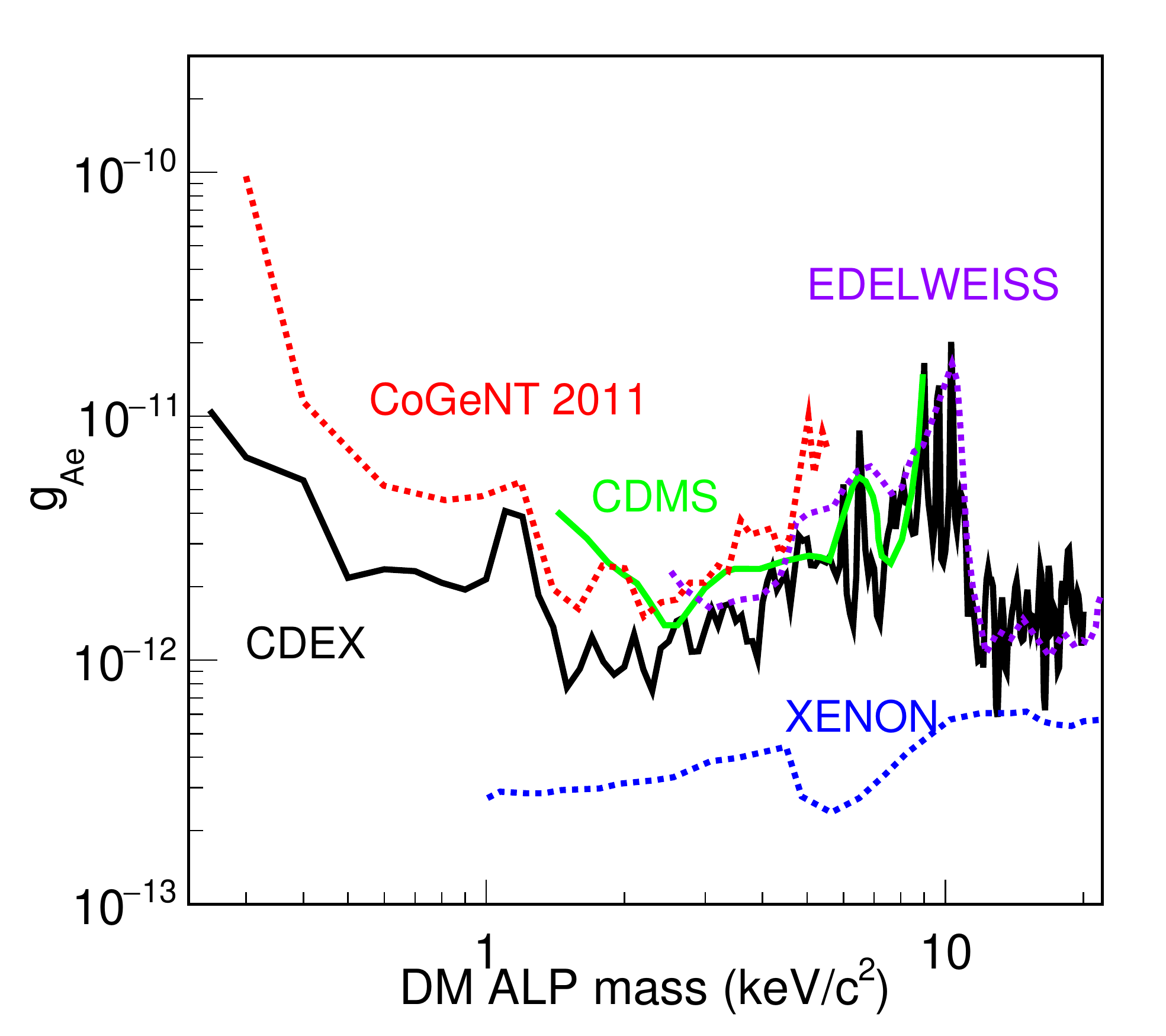}}
  \end{center}
  \caption{Regions in couplings versus the weakly interacting massive
    particle (WIMP) mass parameter space probed and excluded by the
    CDEX-1 experiment with 335.6~kg$\cdot$day of exposure, along with
    comparisons with other benchmark results.  (\textit{a})
    Spin-independent $\chi N$ couplings~\cite{Zhao:2016pr}.
    (\textit{b}) Spin-dependent $\chi n$ couplings~\cite{Zhao:2016pr}.
    (\textit{c}) Axion--electron couplings with solar
    axions~\cite{CDEX:2016ar}.  (\textit{d}) Axion--electron couplings
    with dark matter axions~\cite{CDEX:2016ar}.}
  \label{fig::cdex-explot}
\end{figure}

Figure~\ref{fig::cdex-explot}{\it a} shows the dark matter constraints
on $\chi N$ spin-independent cross sections, along with other selected
benchmark results~\cite{DAMA:2013ep, Savage:2009jcap, CoGeNT:2013pr,
  CDMS:2013pr, S_CDMS:2014pr, CDMS:2016pr, CRESST:2016epII,
  LUX:2014pr}. In particular, the allowed region from the
CoGeNT~experiment \cite{CoGeNT:2013pr} was probed and excluded by the
CDEX-1 results. The anomalous excess originated from the leakage of
surface events into the bulk signal
samples~\cite{Li:2014ap}. Figure~\ref{fig::cdex-explot}{\it b} shows
the constraints on $\chi n$ spin-dependent cross sections, which also
represent an improvement over published results in the light-WIMP
region~\cite{LUX:2016pr,CDMS:2011prII,DAMA:2013ep}.

The superb energy resolution of germanium detectors allows spectral
structures to be resolved even at low energies, where atomic-physics
effects can be important. This benefits studies of axion-like
particles~\cite{Kim:2010rm} because these particles' experimental
signatures are usually manifested as peaks or steps in their energy
depositions at the target. An analysis was performed with the CDEX-1
data, and constraints on axion--electron couplings ($g_{\rm Ae}$) for
solar (Figure~\ref{fig::cdex-explot}{\it c})\ and dark matter axions
(Figure~\ref{fig::cdex-explot}{\it d}) were
derived~\cite{CDEX:2016ar}. In particular, the data represent an
improvement over earlier
results~\cite{EDELWEISS:2013jc,XENON100:2014pr} for dark matter axions
with mass below 1 keV.

The CDEX-1 data set is currently being analyzed with a year-long
exposure.  Studies of annual modulation effects, as well as other
physics channels, are under way. New data are being taken with an
upgraded pPCGe with a lower threshold.

\subsection{Current Efforts and Future Goals}

\begin{figure}[ht]
\begin{center}
  {\bf (a)}\\
  \includegraphics[width=0.52\linewidth]{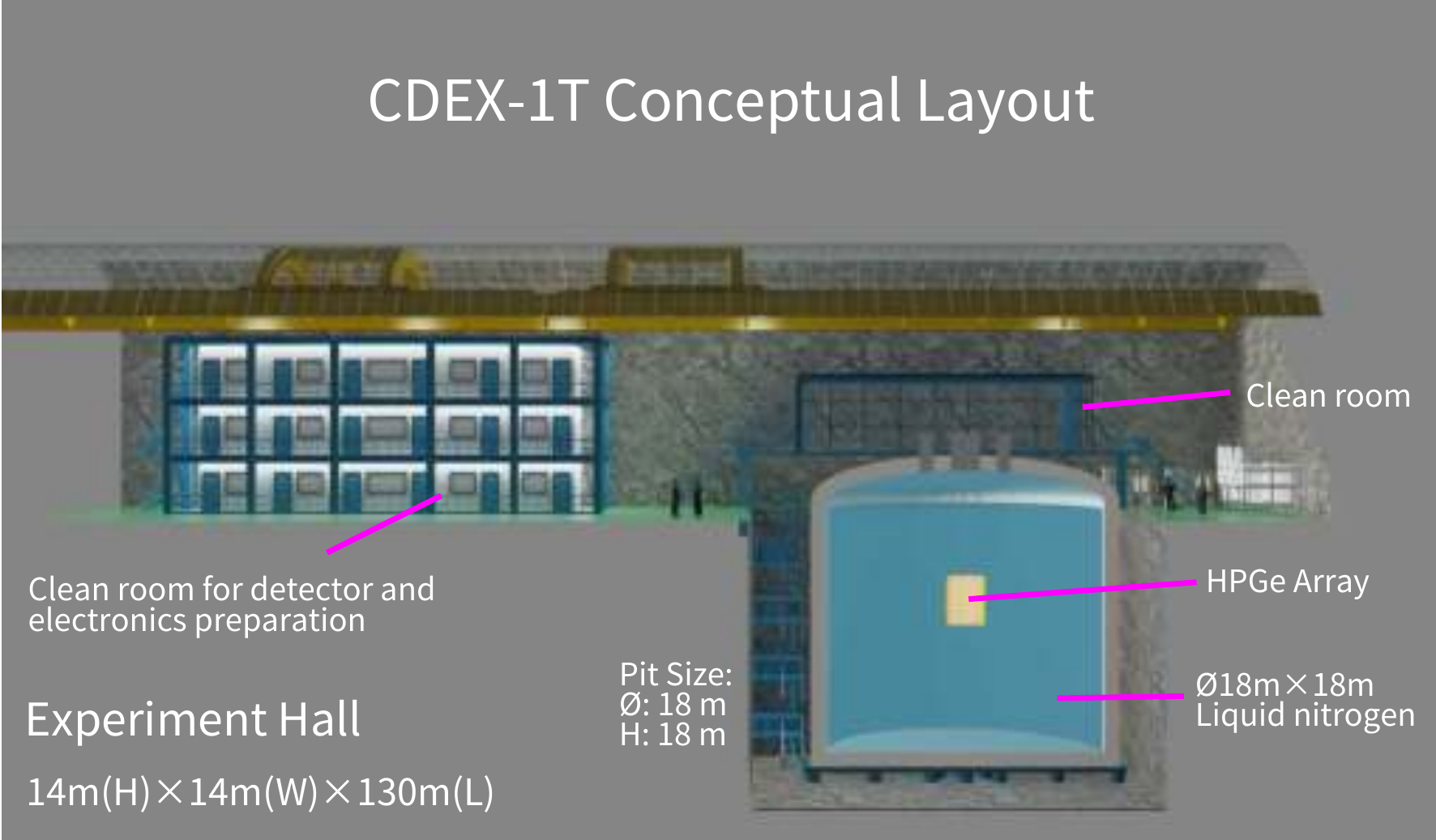} \\
  {\bf (b)}\\
  \includegraphics[width=0.52\linewidth]{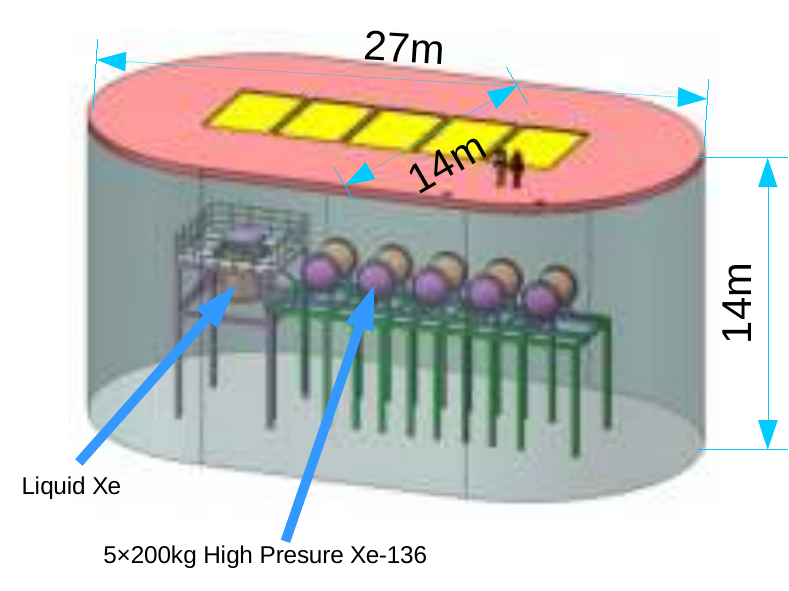} \\
  {\bf (c)}\\
  \includegraphics[width=0.52\linewidth]{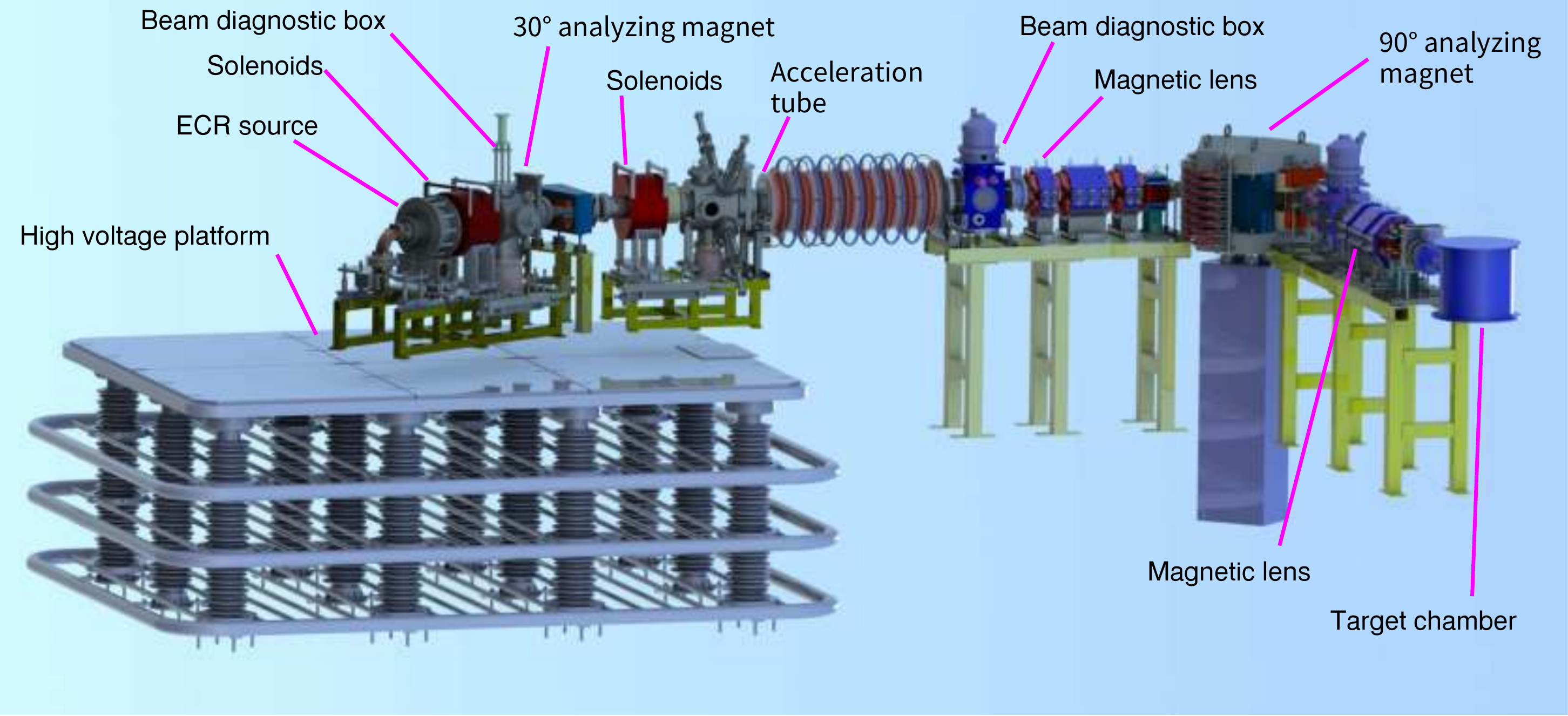}
  \caption{Schematic and conceptual layout of the three ongoing
    research programs expected to be installed at CJPL-II:
    (\textit{a}) CDEX, (\textit{b}) PandaX, and (\textit{c}) JUNA.
    (\textit{a}) CDEX and (\textit{b}) PandaX conduct searches for
    dark matter and neutrinoless double-$\beta$ decay in Hall C and
    Hall B, respectively.  (\textit{c}) JUNA focuses on low-counting
    accelerator-based nuclear physics and astrophysics measurements in
    Hall A.}
  \label{fig::cjpl2-projects}
\end{center}
\end{figure}

The long-term goal of the CDEX program is a ton-scale germanium
experiment (CDEX-1T) searching for dark matter and neutrinoless
double-$\beta$ decay (\znu2b{})~\cite[pp.698--703]{RPP2014}. A pit
with a diameter of 18 m and a height of 18 m has been built in Hall~C
of CJPL Phase~II (CJPL-II) to house such an experiment
(Figure~\ref{fig::cjpl2-projects}{\it a}).

Toward this end, the so-called CDEX-10 prototype has been constructed
with detectors in an array structure with a target mass in the 10-kg
range. CDEX-10 will provide a platform to study the many issues
involving scaling up in detector mass and in improvement of background
and threshold. The detector array is shielded and cooled by liquid
nitrogen (liquid argon may be investigated in the future, as it may
offer an additional benefit of active shielding as AC detector).

Researchers are also pursuing other crucial technology acquisition
projects that would make a ton-scale germanium experiment realistic
and feasible. These include (\textit{a}) detector-grade germanium
crystal growth; (\textit{b}) germanium detector fabrication;
(\textit{c}) development of an ultralow-background, low-temperature,
low-noise preamplifier; and (\textit{d}) production of electroformed
copper, foreseen to be eventually performed underground at CJPL.

The first detector fabricated from commercial crystal that matches
expected performance will be installed at CJPL in 2017. This detector
enables control of assembly materials placed in its vicinity that are
known to be the dominant source of radioactive background, as well as
efficient testing of novel electronics and readout schemes. The
benchmark would be the ability to perform light-WIMP searches with
germanium detectors having \znu2b{}-grade background control. This
configuration would enable detailed studies of potential cosmogenic
tritium contamination in germanium detectors, which in turn would
allow the exploration of strategies to suppress such background.

The projected sensitivities to $\chi N$ spin-independent
(Figure~\ref{fig::cdex-explot}{\it a}) and spin-dependent cross
sections (Figure~\ref{fig::cdex-explot}{\it b}) for CDEX-1T at
targeted specifications in detector threshold, background levels and
data size. The goal of the \znu2b{} program will be to achieve
sufficient sensitivities to completely cover the inverted neutrino
mass hierarchy.  A first \znu2b analysis on the CDEX-1 data was
performed, where an upper limit of
$\tau_{\frac{1}{2}} > 6.4 \times 10^{22} ~{\rm yr}$ at 90\% confidence
level was derived~\cite{cdex1-0nubb}.

\section{PandaX }
\label{sect::pandax}

The PandaX program \cite{CJPL:2010,CJPL:2015} uses xenon as the target
and detector medium in searches for WIMP dark matter, as well as
\znu2b{} in $^{136}$Xe isotopes.  The program follows a staged
plan. The first and second stages (PandaX-I and -II) focus on WIMP
detection at CJPL-I~\cite{CJPL:2010,CJPL:2015}, whereas the future
PandaX-III (a \znu2b{} experiment) and -IV (a multi-ton xenon dark
matter experiment) will be installed in CJPL-II~\cite{CJPL:2015}.

\subsection{Dark Matter Searches with PandaX-I and PandaX-II}

Both PandaX-I and -II utilize a so-called dual-phase xenon
time-projection chamber (TPC) to study recoil signals from xenon
nuclei upon interaction with WIMPs. Two types of signals, S1 and S2,
are recorded for every event. They are produced by scintillation
photons and ionization electrons, respectively. The
XENON~\cite{XENON100:2011pr,XENON100:2012pr},
ZEPLIN~\cite{ZEPLIN:2012pl}, and LUX~\cite{LUX:2014pr} experiments
have demonstrated that the liquid xenon detector techniques offer
exceptional potentials for background suppression and discrimination,
as well as scalability to high target mass, leading the WIMP detection
sensitivity in a wide mass range.

\begin{figure}[ht]
  \begin{center}
    \subfloat[]{\includegraphics[width=.54\textwidth]{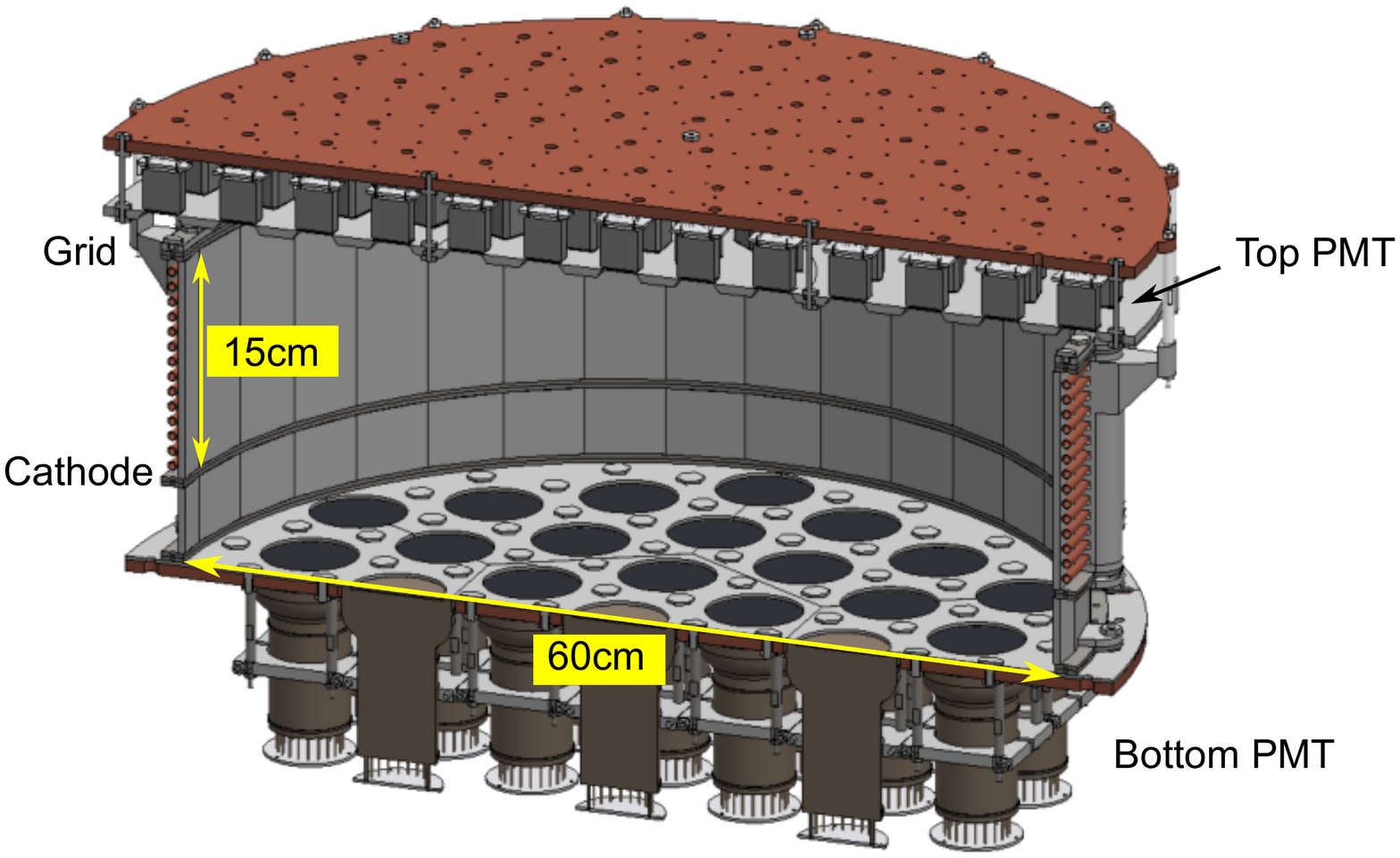}}
    \subfloat[]{\includegraphics[width=.40\textwidth]{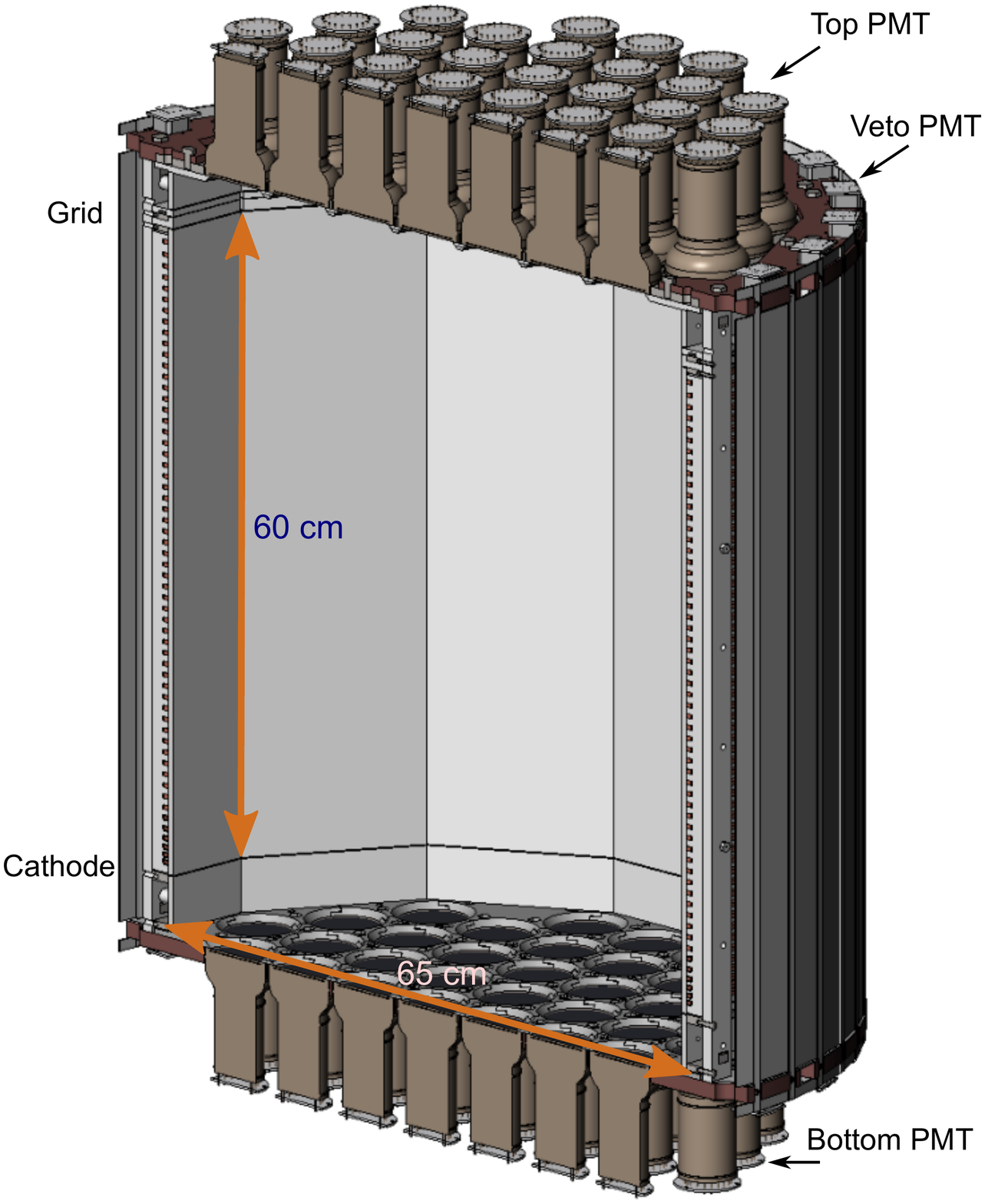}}
  \end{center}
  \caption{
The basic features of the PandaX-I and PandaX-II experiments.
    (\textit{a}) The time-projection chamber (TPC) for PandaX-I, with
    a small drift length of 15~cm between the cathode and anode grid.
    (\textit{b}) The TPC for PandaX-II, with a significantly greater
    drift length of 60~cm. Abbreviation:\ PMT, photo-multiplier
    tube.
}
  \label{fig::pandax::schema}
\end{figure}

PandaX-I and -II share a common infrastructure
(Figure~\ref{fig::pandax::schema})\ (described in detail in Reference
\cite{PANDAX:2014ax}). A passive shield consisting of polyethylene,
lead, and copper was constructed to suppress ambient neutrons and
$\gamma$-rays. The innermost shield is a vacuum copper vessel that is
also a vacuum jacket for the cryostat, as well as a radon barrier. The
cryogenic and gas handling systems~\cite{Gong:2013ji} store,
circulate, and cool the xenon, and have enough capacity for a
ton-scale experiment.

PandaX-I~(Figure~\ref{fig::pandax::schema}{\it a})\
\cite{PANDAX:2014ax,PANDAX:2014cp} was a 120-kg liquid xenon TPC with
polytetrafluoroethylene (PTFE) reflective walls to enhance light
collection. Its top and bottom photo-multiplier tube (PMT) arrays
consisted of, respectively, 143 and 37 PMTs with diameters of 1 and 3
inches that collected the S1 and S2 photons from events originating in
the active volume. The upgrade to PandaX-II
(Figure~\ref{fig::pandax::schema}{\it b}) includes a new
low-background steel inner vessel, an increased liquid xenon target
mass of 580 kg, and a total of 110 3-inch-diameter PMTs. An additional
48 1-inch-diameter PMTs serve as an active veto in the optically
isolated 4-cm liquid xenon region surrounding the outer circumference
of the TPC, providing extra power for background
rejection~\cite{PANDAX:2016pr}.

\begin{figure}[ht]
  \begin{center}
    \subfloat[]{\includegraphics[width=.5\linewidth]{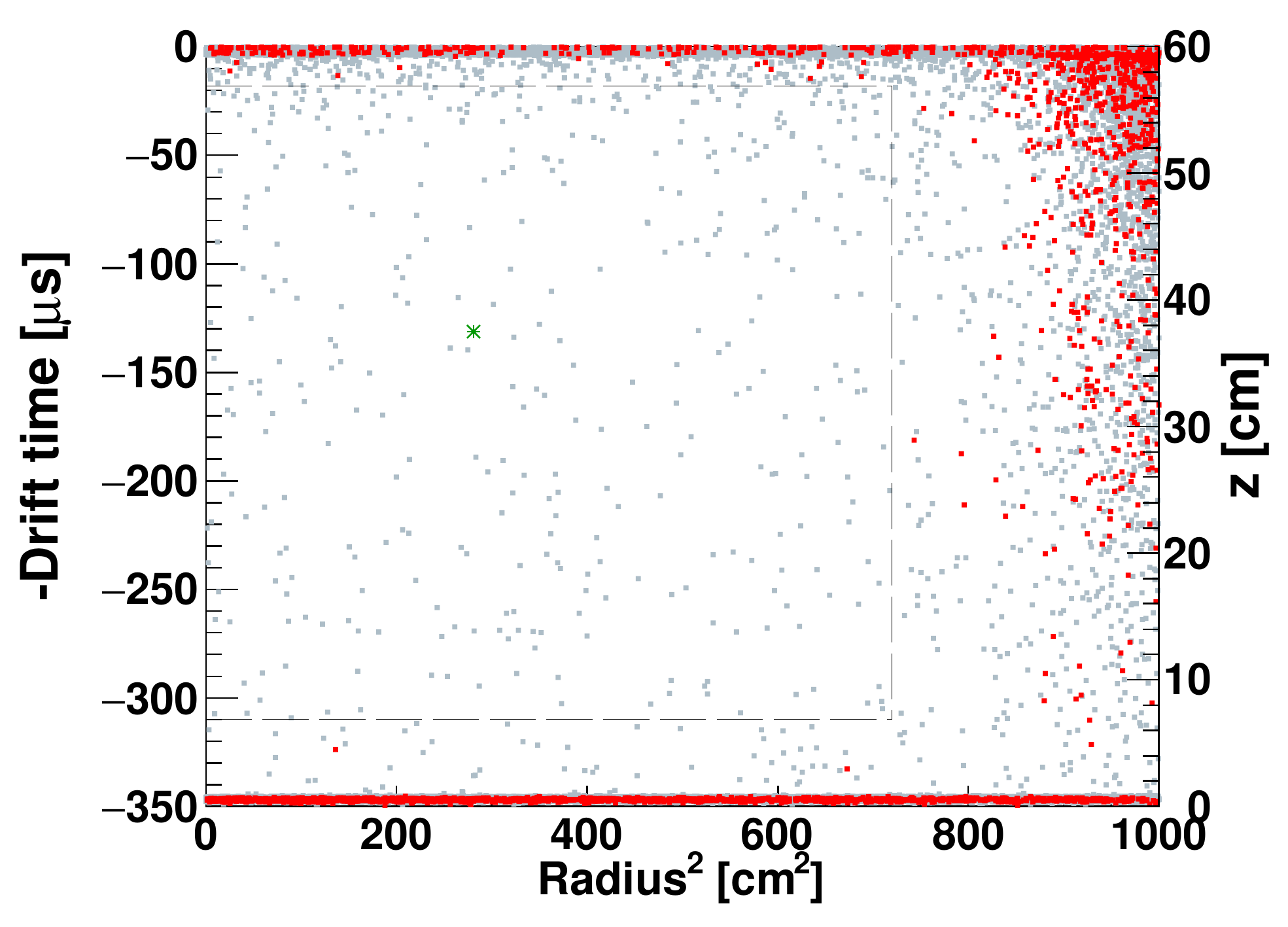}}
    \subfloat[]{\includegraphics[width=.5\linewidth]{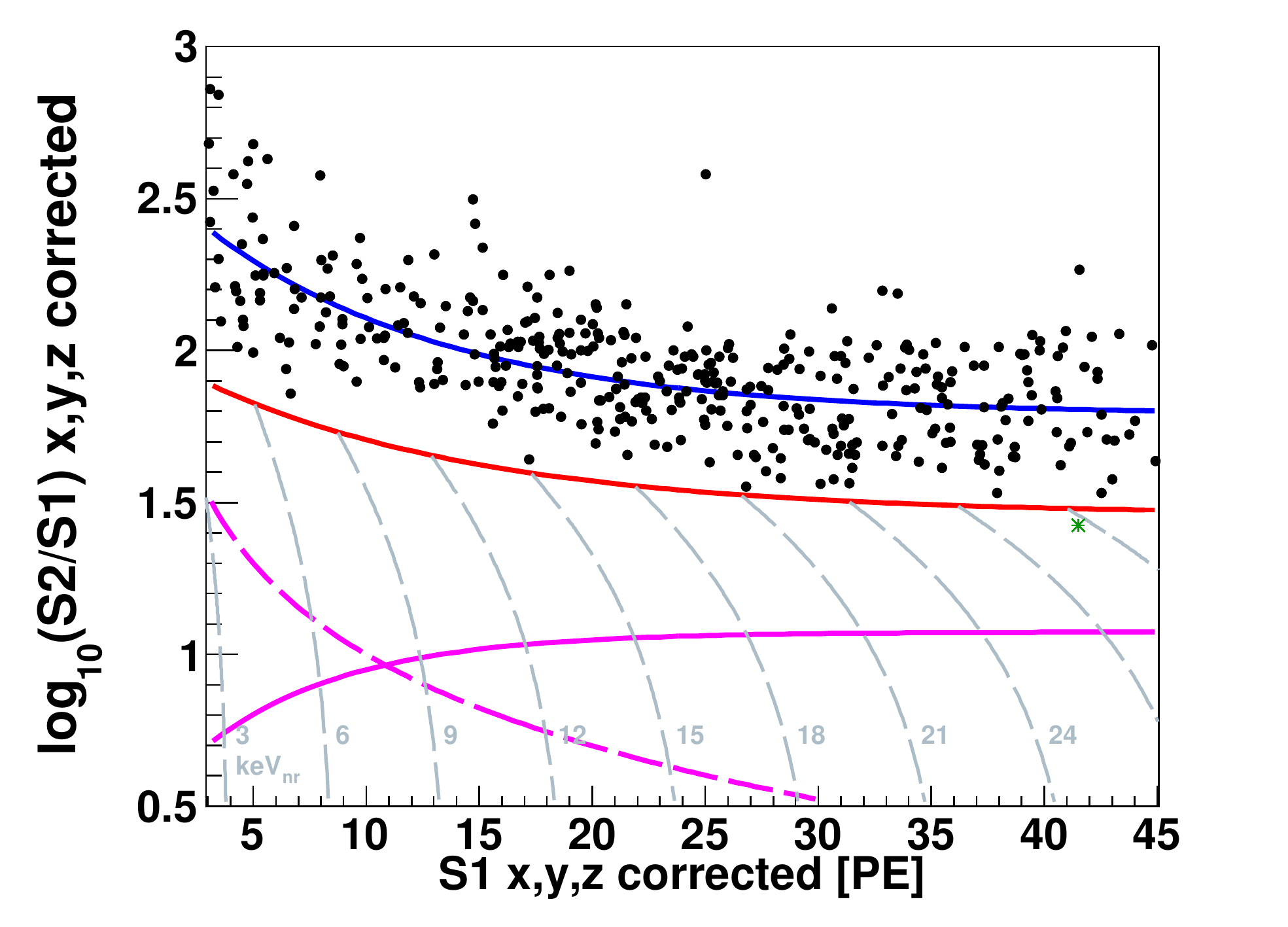}}
  \end{center}
  \caption{Event distributions from the PandaX-II data.  (\textit{a})
    Spatial distribution of the events and the defined fiducial volume
    to select candidate events.  (\textit{b}) The band of
    $\log_{10}{\text{S2}/\text{S1}}$ versus S1 from PandaX-II events
    within the fiducial volume. The red solid line represents the
    fitted median derived from nuclear recoil calibration data. Weakly
    interacting massive particle candidate signals are those below the
    median line; no such events were observed.}
  \label{fig::pandax_ii::dis_events}
\end{figure}

Candidate events from $\chi N$ elastic scattering are selected
according to their locations in the central fiducial volume
(Figure~\ref{fig::pandax_ii::dis_events}{\it a}). In addition,
correlations of the S1 and S2 signatures allow differentiation of
nuclear recoil events versus electron recoil
background. Figure~\ref{fig::pandax_ii::dis_events}{\it b} shows the
$\log_{10}{\text{S2}/\text{S1}}$ band of potential signal events,
along with calibration data with a neutron source.

\begin{figure}[ht]
  \begin{center}
    \subfloat[]{\includegraphics[width=0.48\textwidth]{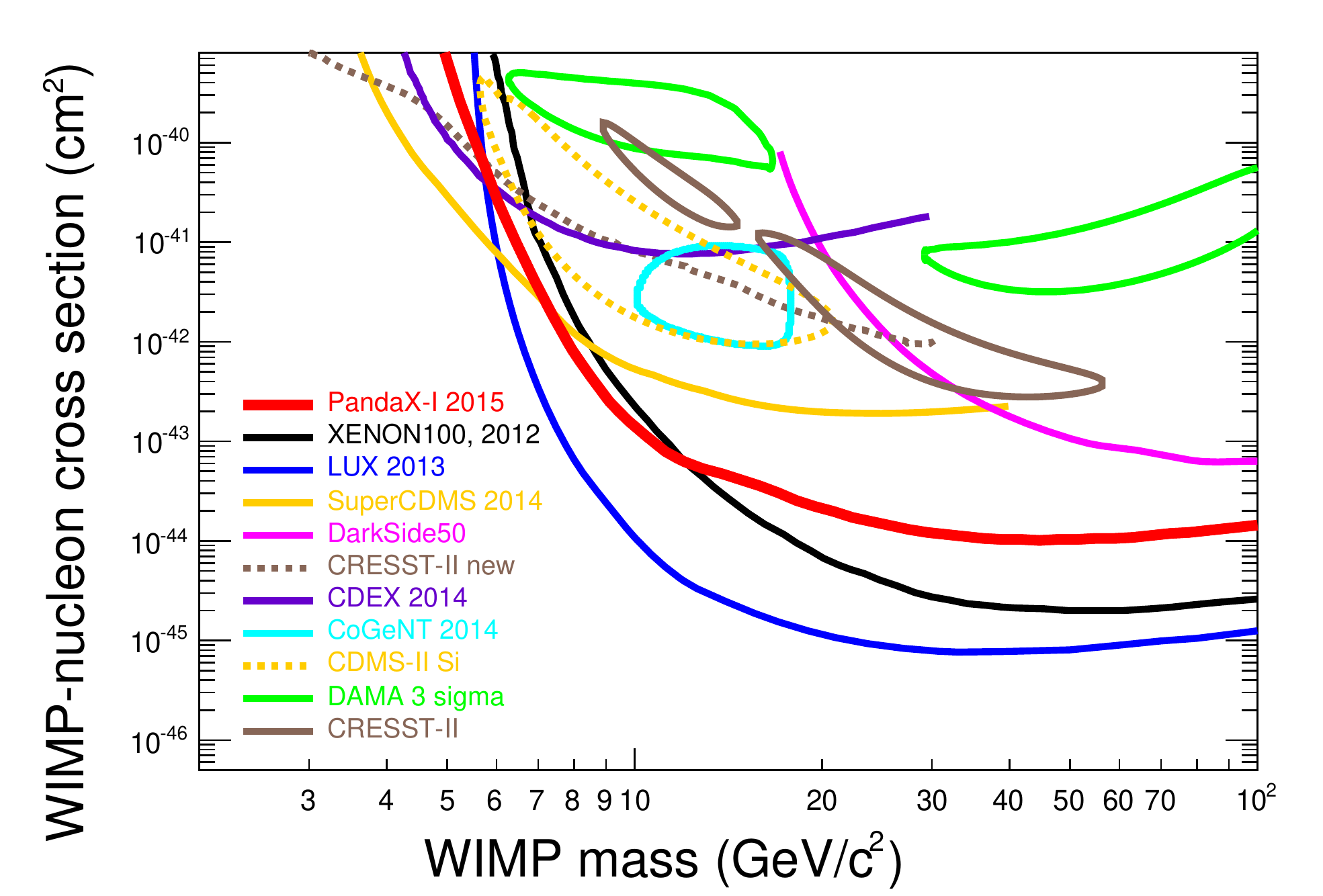}}
    \subfloat[]{\includegraphics[width=0.48\textwidth]{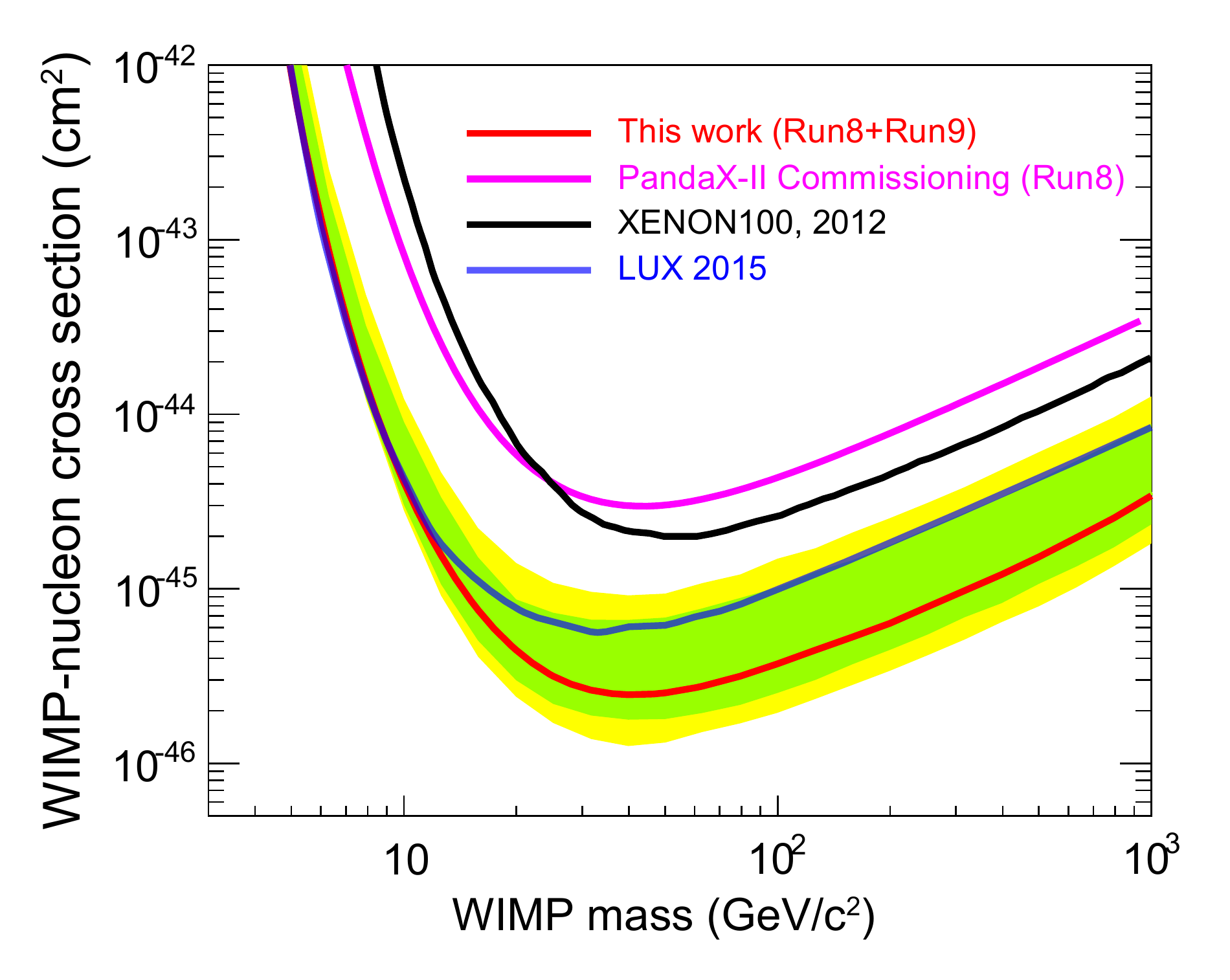}}
    \newline
    \subfloat[]{\includegraphics[width=0.49\linewidth]{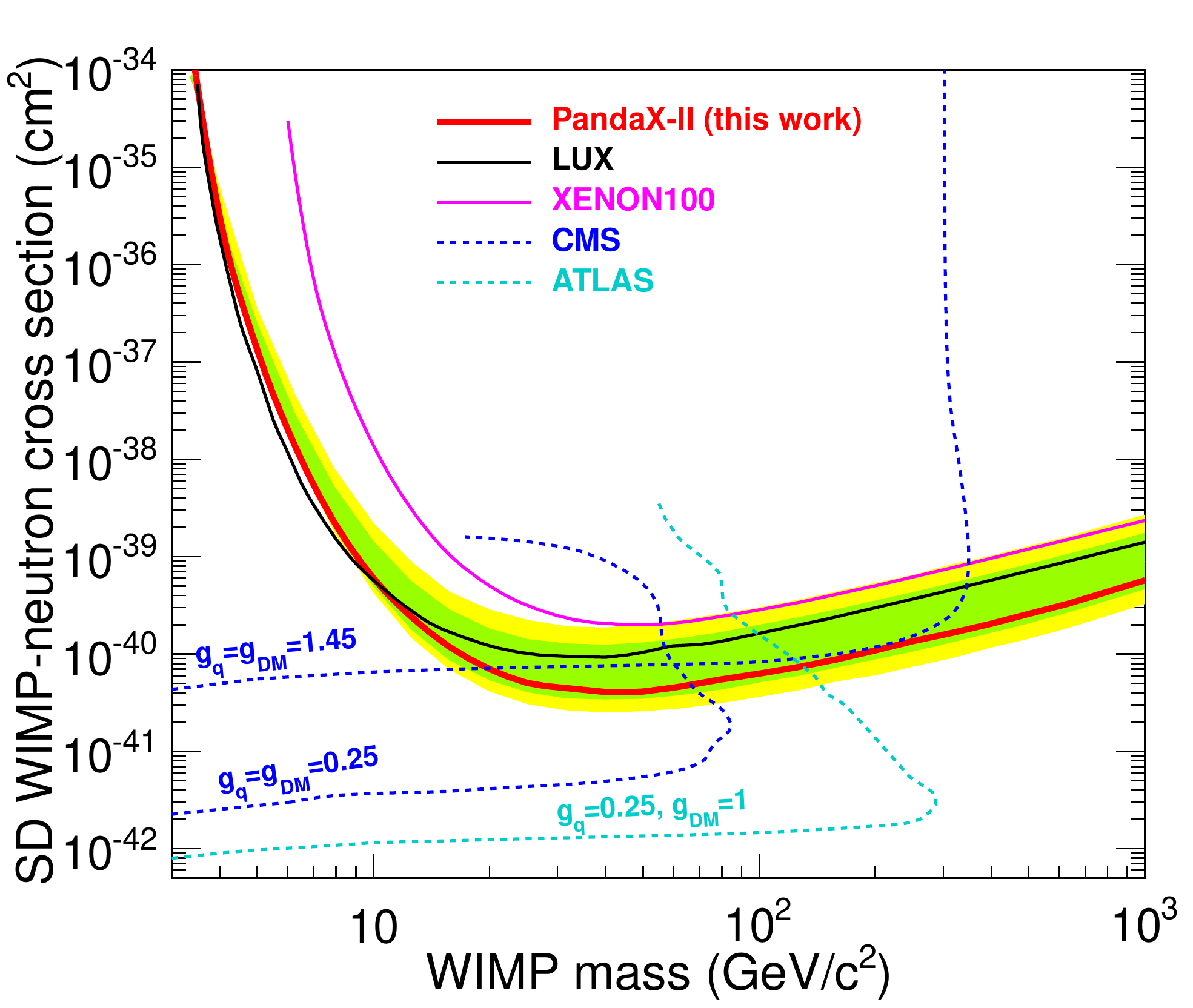}}
    \subfloat[]{\includegraphics[width=0.49\linewidth]{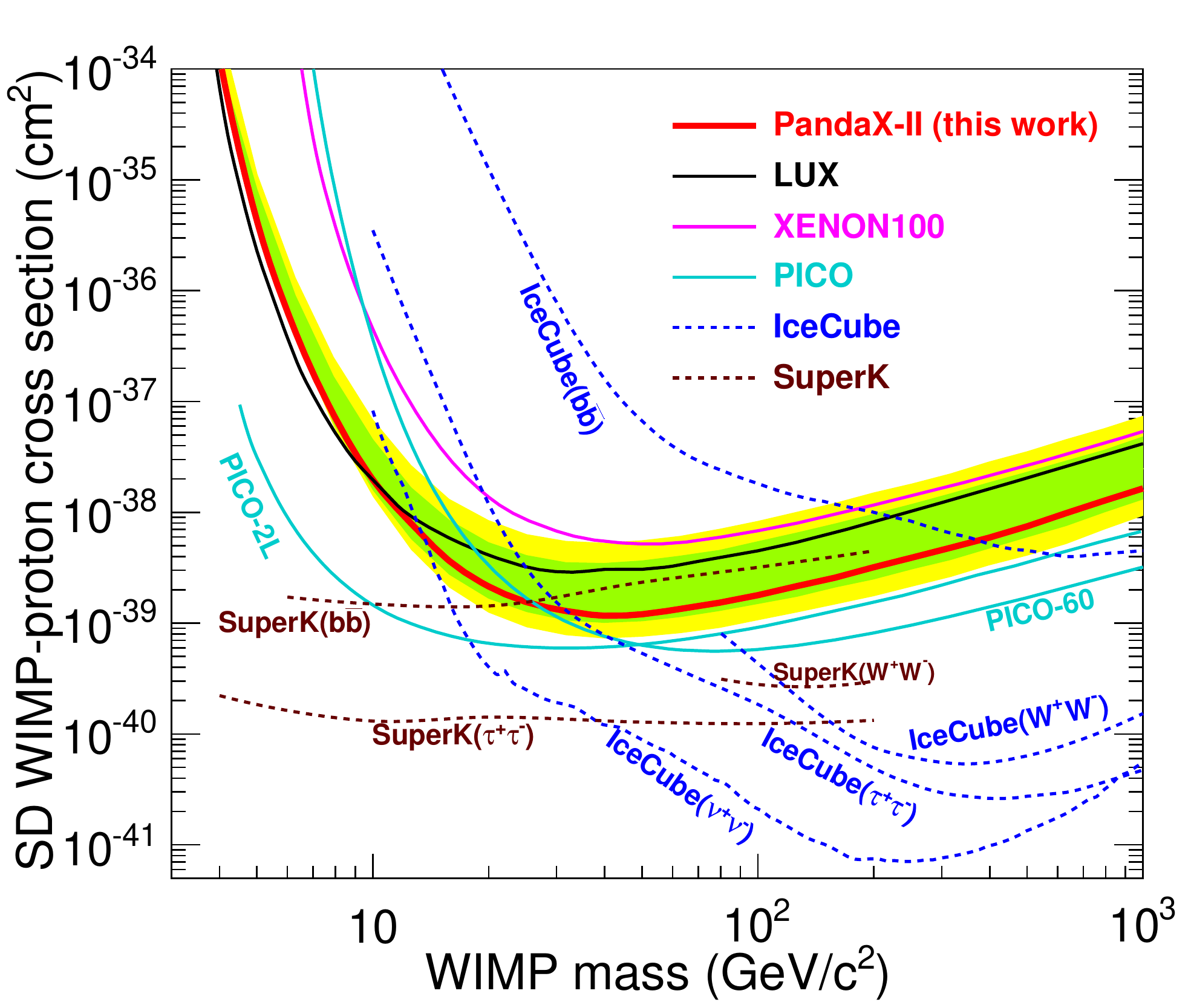}}
  \end{center}
  \caption{Upper limits at the 90\% confidence level (\textit{solid
      red lines}) set by the PandaX experiments for the $\chi N$ cross
    sections.  (\textit{a}) Limits on the spin-independent $\chi N$
    scattering cross section set by the PandaX-I full exposure
    data~\cite{PANDAX:2015pr}, compared with the world data.
    (\textit{b}) Limits on the spin-independent $\chi N$ scattering
    cross section set by the PandaX-II data with
    $3.3 \times 10^{4}$~kg$\cdot$day exposure in 98.7 live
    days~\cite{PANDAX:2016pl}.  (\textit{c}) Limits on the
    spin-dependent WIMP--neutron scattering cross section set by the
    same PandaX-II data set as in panel \textit{b~}\cite{Fu:2016ega}.
    (\textit{d}) Limits on the spin-dependent WIMP--proton scattering
    cross section set by the same PandaX-II data set as in panel
    \textit{b}~\cite{Fu:2016ega}. Abbreviation:\ WIMP, weakly
    interacting massive particle.  }
  \label{fig::pandax::data}
\end{figure}

The PandaX-I experiment collected data for dark matter searches
between May and October 2014. The final results from PandaX-I are
presented in Reference~\cite{PANDAX:2015pr} with a total exposure of
$54 \times 80.1$ kg$\cdot$day exposure. The data distribution was
consistent with the background-only hypothesis. A profile likelihood
analysis was performed with all candidate events and the expected WIMP
and background distributions, leading to the exclusion limit of the
$\chi N$ spin-independent elastic scattering cross section shown in
Figure~\ref{fig::pandax::data}{\it a}. At the 90\% confidence level,
the results are incompatible with the spin-independent isoscalar WIMP
interpretation of all previously reported positive signals from the
DAMA/LIBRA~\cite{DAMA:2008ep,DAMA:2010ep,DAMA:2013ep},
CoGeNT~\cite{CoGeNT:2011pr,CoGeNT:2013pr},
CRESST-II~\cite{CRESST:2012ep}, and CDMS-II-Si~\cite{CDMS:2013pr}
experiments.  The bounds on the $\chi N$ cross section are more
stringent than those from SuperCDMS~\cite{S_CDMS:2014pr} above a WIMP
mass of 7 GeV/\textit{c}$^2$. Below a mass of 5.5 GeV/\textit{c}$^2$,
the results were the tightest reported constraints among all
xenon-based experiments.

The upgrade from PandaX-I to PandaX-II was completed in 2015. The
physics commissioning run performed from November to December 2015
obtained a dark matter exposure of $306 \times 19.1$ kg$\cdot$day.
The data taking concluded with the identification of a strong
$^{85}$Kr background. The analysis of this data set has been published
\cite{PANDAX:2016pr}. No WIMP candidates were identified, and a 90\%
upper limit comparable to the 225-day limit from
XENON100~\cite{XENON100:2012pr} was set. PandaX-II resumed physics
data taking in March 2016 after a krypton distillation campaign that
reduced the krypton level by a factor of 10, leading to an average
electron-recoil background of $2\times 10^{-3}$
events/(kg$\cdot$day)---the world's lowest reported level. The
combined data from the commissioning run and an additional data-taking
period from March to June 2016 provide a total exposure of
$3.3 \times 10^{4}$~kg$\cdot$day~\cite{PANDAX:2016pl} and represent
the largest WIMP search data set to date.

After the data-quality cuts, the S1 and S2 signal range cut, and the
boosted decision tree cut were applied to the collected data, 389
candidates in the fiducial volume were selected from among
$2.45\times 10^{7}$ events
(Figure~\ref{fig::pandax_ii::dis_events}{\it a}). No WIMP candidates
were observed (Figure~\ref{fig::pandax_ii::dis_events}{\it b}).

On the basis of a profile likelihood analysis of all data, an
exclusion limit on the spin-independent $\chi N$ cross section was
obtained (Figure~\ref{fig::pandax::data}{\it b}). The most stringent
limit, $2.5\times10^{-46}$ cm$^2$ at a WIMP mass of 40
GeV/\textit{c}$^2$, represents an improvement of more than a factor of
10 from the limit in Reference~\cite{PANDAX:2016pr}. In the
high-WIMP-mass region, the bounds improve by more than a factor of two
over the previous most stringent limit~from LUX \cite{LUX:2016pl}.

By using the same data set, PandaX-II set new constraints on the
spin-dependent $\chi n$ (Figure~\ref{fig::pandax::data}{\it c}) and
$\chi p$ (Figure~\ref{fig::pandax::data}{\it d}) cross
sections. Compared with other direct detection experiments, PandaX-II
has derived the most stringent upper limits on the $\chi N$ cross
sections for WIMPs with masses above 10 GeV/\textit{c}$^2$.

\subsection{Current Efforts and Future Goals}

PandaX-II is anticipated to continue physics data taking until 2018,
increasing the exposure by a factor of five. In parallel, a future
PandaX-xT dark matter experiment and a PandaX-III \cite{Chen:2016qcd}
\znu2b{} experiment are being planned. Both experiments will be
located in Hall~B of CJPL-II, in which a large, ultrapure water pool
will be used as shielding for both
experiments. Figure~\ref{fig::cjpl2-projects}{\it b} shows the
conceptual layout of the planned experiments.

The PandaX-xT experiment will be a multi-ton dark matter experiment
employing the same dual-phase xenon TPC technology as PandaX-I and
-I. Its ultimate goal is to reach the so-called neutrino floor at a
sensitivity of $\sim10^{-49}$ cm$^2$ for the spin-independent $\chi N$
cross section at a WIMP mass of $\sim100$ GeV/\textit{c}$^2$. The plan
of the initial stage of PandaX-xT is to construct a detector with a
sensitive target mass of 4 tons of liquid xenon. The TPC will be of a
cylindrical shape roughly 1.2 m in diameter and height, almost a
factor of two larger in every dimension in comparison to PandaX-II. For the
photo-sensor coverage, it will use in total more than 350 3-inch PMTs
at the top and bottom. The powerful self-shielding from liquid xenon
will significantly suppress detector-material-related background. The
level of $^{85}$Kr background is also expected to reduce greatly by
performing online distillation during the detector operation. The
total background rate is projected to be 0.05$\times$10$^{-3}$
evts/kg/day, a factor of 40 lower than the best level achieved in
PandaX-II.  With a 2-year exposure, the sensitivity to the $\chi N$
cross section is expected to reach $10^{-47}$ cm$^2$, more than one
order of magnitude lower than the current PandaX-II limit.  Increasing
detector size does increase the technological complexity. For example,
larger electrodes require new ways to keep their rigidity and flatness
when the electric field is applied, while maintaining the
thin-wire-plane to ensure high photon detection efficiency. As another
example, more readout channels and longer readout windows are
required, demanding a more sophisticated data acquisition system
capable to handle the increasing data bandwidth. Therefore, the 4-ton
experiment will also serve to demonstrate the critical technology
necessary for the future-stage PandaX-xT detector.

The PandaX-III experiment will search for \znu2b{} events using
high-pressure gas detectors with enriched $^{136}$Xe. In the first
phase of PandaX-III, a gaseous TPC with 200 kg of 90\%-enriched
$^{136}$Xe will operate at 10-bar pressure, with a charge readout
plane made of microbulk micromesh gaseous structure (micromegas)
detectors.  It will be the first 100-kg stage \znu2b{} experiment
hosted in China.  In addition to having good energy and spatial
resolution, the TPC will be able to reconstruct tracks from \znu2b{}
events, leading to powerful background suppression.

With a single 200-kg module, the half-life sensitivity to $^{136}$Xe
\znu2b{} decay will reach $10^{26}$ years after 3 years of
operation. With multiple modules with improved background levels and
energy resolution, this program could be upgraded to a ton-scale
experiment with a half-life sensitivity of $10^{27}$ years, which
would enable the discovery of \znu2b{} if neutrinos are Majorana
fermions with the inverted mass ordering.

\section{LOW-BACKGROUND FACILITIES}
\label{sect::lobkgf}

Table~\ref{tab::cjpl_bkg} summarizes measurements of ambient radiation
background by the main experiment hall in CJPL-I. Standard detector
techniques~\cite{Wu:2013cp,Zeng:2014jr,Mi:2015cs,Zeng:2015ni} were
adopted to perform these measurements.

The residual cosmic-ray flux at CJPL is $(2.0\pm{}0.4)\times 10^{-10}$
cm$^{-2}$~s$^{-2}$ or, equivalently, $\sim$5 m$^{-2}$ month$^{-1}$,
measured with two groups of triple-layer plastic scintillator
telescopes~\cite{Wu:2013cp}. Owing to the large rock overburden, the
attenuation factor of $10^{-8}$ relative to the typical cosmic-ray
fluxes on the surface is the largest of all operational underground
laboratories (Figure~\ref{fig::cjpl_mflux}).

%%  ======== Table 2 ===========

\begin{table}[h]
  \caption{Summary of the environmental radiation background measurements at the
    main experiment hall of China Jinping Underground Laboratory, Phase I}
  \label{tab::cjpl_bkg}
\begin{center}
  \begin{tabular}{lll}
    \hline
    {\bf Background radioactivity} & {\bf Detection techniques} & {\bf Measurement results} \\
    \hline
    Cosmic-muon flux~\cite{Wu:2013cp} & Plastic scintillator telescope & $(2.0 \pm 0.4) \times 10^{-10}$ cm$^{-2}$ s$^{-1}$\\
    \hline
    Radioactivity of bedrock~\cite{Zeng:2014jr} & In situ high-purity & \\
    $^{238}$U series & ~~germanium detector & 3.69--4.21 Bq kg$^{-1}$\\
    $^{232}$Th series & & 0.52--0.64 Bq kg$^{-1}$\\
    $^{40}$K & & 4.28 Bq kg$^{-1}$ \\
    \hline
    Radioactivity of concrete~\cite{Zeng:2014jr} & In situ high-purity & \\
    $^{238}$U series & ~~germanium detector & \\
    $^{214}$Pb & & 19.88 Bq kg$^{-1}$ \\
    $^{214}$Bi & & 16.03 Bq kg$^{-1}$ \\
    $^{232}$Th series& & \\
    $^{228}$Ac & & 7.38 Bq kg$^{-1}$ \\
    $^{212}$Pb & & 7.48 Bq kg$^{-1}$ \\
    $^{208}$Tl & & 8.15 Bq kg$^{-1}$ \\
    $^{40}$K & & 36.67 Bq kg$^{-1}$ \\
    \hline
    Air absorbed dose rate & Ionization chamber &  \\
    Main hall & & 19.27 nGy h$^{-1}$\\
    Inside polyethylene room  & & 0.43 nGy h$^{-1}$\\
    \hline
    Air radon concentration & Ionization chamber & \\
    December 2010--September~2011~\cite{Mi:2015cs} & ~~ for $\alpha$ counting & \\
    Unventilated & & $101 \pm 14$ Bq m$^{-3}$\\
    Ventilated & & $86 \pm 25$ Bq m$^{-3}$\\
    January~2015--December~2015 & & \\
    Unventilated (January--March) & & $108 \pm 50$ Bq m$^{-3}$\\
    Ventilated (April--December) & & $45 \pm 28$ Bq m$^{-3}$\\
    \hline
    Neutron& & \\
    Thermal neutron & $^{3}$He-proportional tube~\cite{Zeng:2015ni} & $(4.00\pm 0.08)\times 10^{-6}$ cm$^{-2}$ s$^{-1}$ \\
                                   & Multiple Bonner spheres~\cite{Hu:2016ax} & $(7.03 \pm 1.81) \times 10^{-6}$ cm$^{-2}$ s$^{-1}$ \\
    Fast neutron & Liquid scintillator & $(1.50 \pm 0.07) \times 10{-7}$ cm$^{-2}$ s$^{-1}$ \\
                                   & Multiple Bonner spheres~\cite{Hu:2016ax} & $(3.63 \pm 2.77) \times 10^{-6}$ cm$^{-2}$ s$^{-1}$ \\
    Total neutron flux & Multiple Bonner spheres~\cite{Hu:2016ax} & $(2.69 \pm 1.02) \times 10^{-5}$ cm$^{-2}$ s$^{-1}$ \\
    \hline
  \end{tabular}
\end{center}
\end{table}

%%  ===================

The $\gamma$ radioactivity from the bedrock, as well as the concrete
components of CJPL-I, were measured with a in situ high-purity
germanium detector~(Table~\ref{tab::cjpl_bkg})\
\cite{Zeng:2014jr}. The bedrock of Jinping Mountain is mainly marble,
providing relatively good radiopurity. The measured radioactivity
levels compare favorably to those of other underground
laboratories~\cite{GRANSASSO:2012ep,SNOLAB:2012ep}.

The materials used to build CJPL-I were adapted from the construction
of the nearby hydro-power station, and no radioactivity screening
procedures were performed. Consequently, the radioactivity levels of
the CJPL-I concrete are higher than those of other operational
underground laboratories. In the construction of CJPL-II (discussed in
Section~\ref{sect::cjpl2}), radioactivity from various concrete
samples have been measured, and only those with low radiopurity were
selected.

The absorbed dose rate in air is 19.27 nGy h$^{-1}$, as determined
with a high-pressure ionization chamber. This dose rate is lower than
typical levels at surface locations. Therefore, low-dose-rate
calibration for radiation environment dosimetry can be performed at
CJPL-I.

The air radon concentration in the main experiment hall at CJPL-I is
monitored with commercial ionization detectors. During the early
operation of CJPL-I from December~2010 to September~2011, the radon
concentration was $101\pm 14$ Bq m$^{-3}$ without ventilation, and
$86\pm 25$ Bq m$^{-3}$ when the ventilation system was fully
operational. In 2015, improvements to the ventilation systems resulted
in a reduced level of $45\pm{}28$ Bq m$^{-3}$ or $108\pm{}50$ Bq
m$^{-3}$ when the ventilation was on or off,
respectively. Occasionally, there was a sudden surge of radon
concentration to several hundred becquerels per cubic meter despite
the ventilation system being fully operational.  The cause of these
surges is unknown and is under investigation.

Neutron backgrounds are measured by multiple detector systems to match
their different energy ranges. A gaseous $^{3}$He proportional
ionization chamber housed in a 100-cm-long tube enables measurements
of a thermal neutron flux of $(4.00\pm{}0.08)\times 10^{-6}$ cm$^{-2}$
s$^{-1}$. A 30-L liquid scintillator loaded with gadolinium tags the
fast neutron background via delayed coincidence of proton recoils
followed by thermal neutron capture, enabling measurements of
$(1.50\pm{}0.07)\times 10^{-7}$ {cm$^{-2}$ s$^{-1}$} in the main hall
of CJPL-I. A multiple Bonner sphere neutron spectrometer, with
$^{3}$He ionization chambers enclosed in polyethylene shielding of
different thickness, is used to measure neutron spectra below 20
MeV. The total flux is $(2.69\pm{}1.02)\times 10^{-5}$ {cm$^{-2}$
  s$^{-1}$}, of which $(7.03\pm{}1.81)\times 10^{-6}$ {cm$^{-2}$
  s$^{-1}$} are thermal (below 0.5 eV) neutrons and
$(3.63\pm{}2.77)\times 10^{-6}$ {cm$^{-2}$ s$^{-1}$} are fast
neutrons.

Neutron measurements are also performed inside the polyethylene
shielding room in the main CJPL-I hall. The thermal neutron flux is
$(1.75\pm{}0.27)\times 10^{-7}$ {cm$^{-2}$ s$^{-1}$}, as determined
with a $^{3}$He proportional ionization chamber, whereas the fast
neutron flux is $(4.27\pm{}0.72)\times 10^{-9}$ {cm$^{-2}$ s$^{-1}$},
as determined with gadolinium-loaded liquid scintillator. These
measurements demonstrate that the 100-cm-thick polyethylene room can
efficiently attenuate the neutron flux.

Efficient low-level counting facilities are essential for underground
experiments. Currently there are two low-background counting
facilities, GeTHU-I and GeTHU-II, which employ high-purity germanium
detectors installed within custom-designed shielding
structures~\cite{Zeng:2014ar}. The best integrated background rate
achieved to date is 0.42 counts min$^{-1}$ (kg$\cdot$Ge)$^{-1}$
between 40 and 2,700 keV. The copper shielding of these germanium
detector systems contains residual radioactivity. Future upgrades
would reduce the background by use of fine radiopure electroformed
copper, instead of commercially produced copper. The GeTHU systems
provide support for the various CJPL\ experiments.

\section{CHINA JINPING UNDERGROUND LABORATORY: PHASE II}
\label{sect::cjpl2}

\begin{figure}[ht]
  \includegraphics[width=0.9\textwidth]{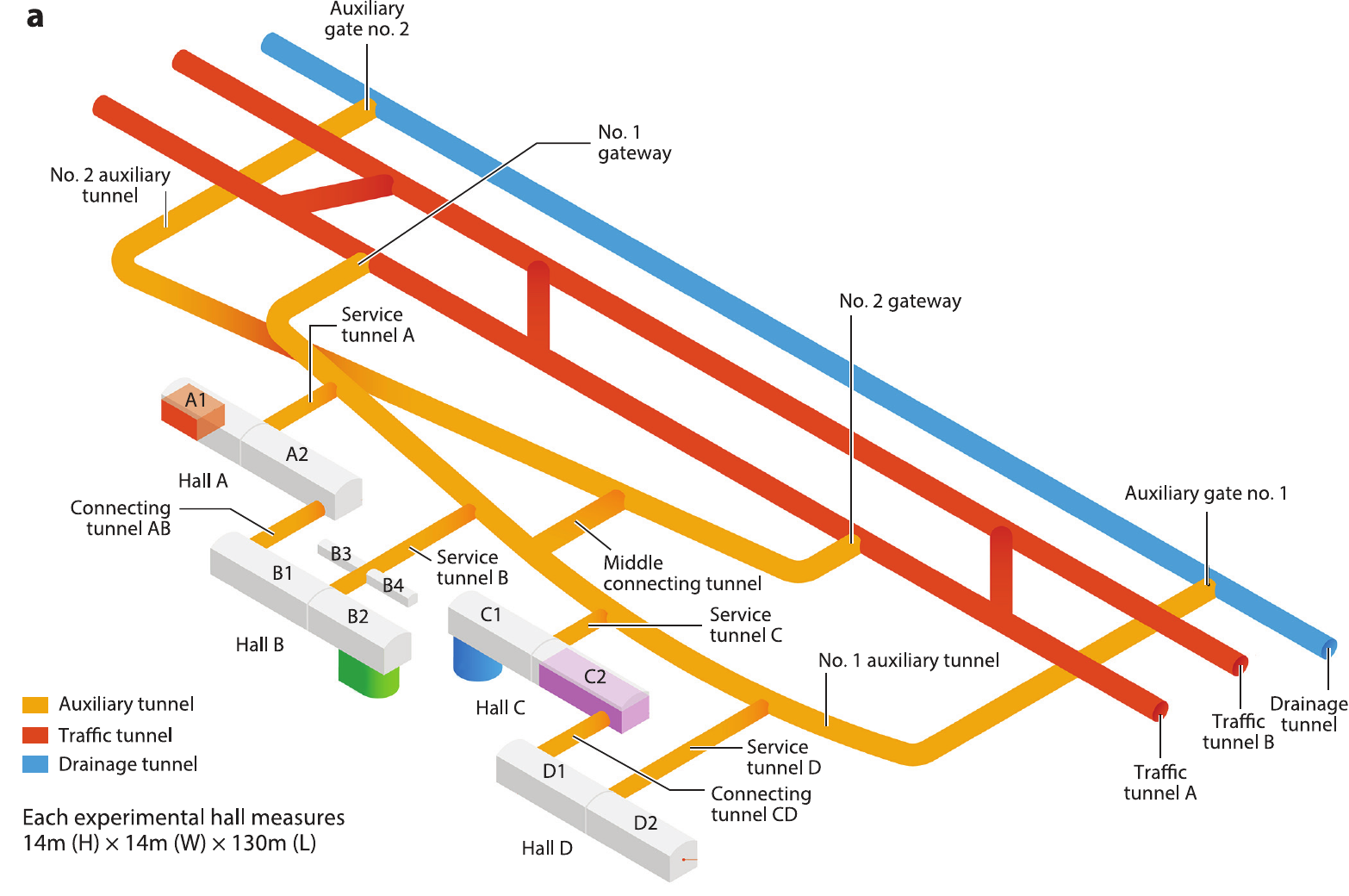}\\
  \includegraphics[width=0.9\textwidth]{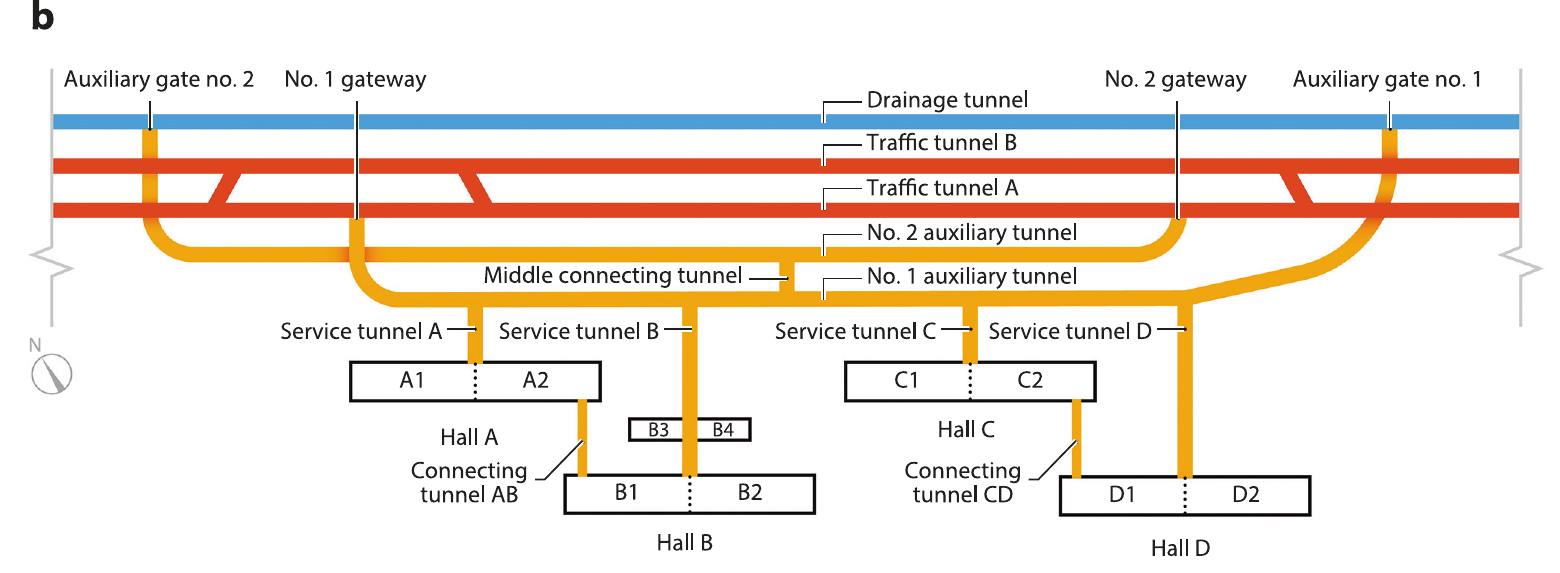}
  \caption{Schematic layout of CJPL-II, showing the four experimental
    halls as well as the connecting and service tunnels.}
  \label{fig::cjpl2_schema}
\end{figure}

The commissioning of CJPL-I and the development of its scientific
programs attracted great interest both domestically and
internationally. Construction on the expansion of the facility began
in 2014. Figure~\ref{fig::cjpl2_schema} depicts a schematic diagram of
the second phase of CJPL\ (CJPL-II). The key dimension parameters
characterizing access and installation of future experiments are
summarized in Table~\ref{tab::cjpl2_access}.

CJPL-II is located 500 m to the west of CJPL-I, along the same road
tunnel. It will have four caverns, each with dimensions of 14 m
(width) $\times$ 14 m (height) $\times$ 130 m (length), and will be
interconnected with access and safety tunnels. The laboratory floor
area will be approximately 20,000~m$^2$.  Two pits will provide
additional headroom for specific applications---one with diameter 18 m
and height 18 m and another with length 27 m, width 16 m, and depth 14
m. The total space, including internal traffic tunnels and common
facilities, will be approximately 300,000~m$^3$.  It is planned that
additional ground laboratory, offices, meeting venues and logistics
support will be built before 2020 to accommodate 150 scientists to
live and work in a campus which is in the neighbourhood of another 600
workers from the hydro-power company.

%% ==========   Table 3 =================

\begin{table}[h]
  \caption{Summary of the key dimensions of CJPL-II
relevant to the access and installation of experimental hardware.
}
  \label{tab::cjpl2_access}
  \begin{center}
    \begin{adjustbox}{max width=\textwidth}
      %% \begin{tabular}{@{}l|c|c|c|c@{}}
      \begin{tabular}{lcl}
        \hline
Items & Dimension & Description \\ \hline \hline
Main  &   Height = 8 m     &  Two Parallel Tunnels;  \\
~~ Road Tunnels  & Width = 8 m
&  ~~ 9.2(8.3)~km from East(West) Entrance \\ \hline
Internal Tunnel & Height = 8 m  &  Connecting One of Main Tunnels \\
 & Width = 8 m   &  ~~ with Experimental Halls \\
&   Length = 800 m  & \\ \hline
Connecting Tunnels & Height = 8 m  &  One for each Hall; \\
 & Width = 8 m   &  ~~ 64~m and 134~m from Internal Tunnel \\ \hline
Experimental Halls &   Height = 14 m  &  Four Halls in total; \\
&   Width = 14 m   &  ~~ Entrances at Middle Sections  \\
&   Length = 130 m  & \\ \hline
Hall B Pit & Length = 27 m  &  \\
& Width = 16 m & \\
& Depth = 14 m & \\ \hline
Hall C Pit & Diameter = 18 m  &  \\
& Depth = 18 m & \\ \hline
      \end{tabular}
    \end{adjustbox}
  \end{center}
\end{table}

%% ====================

%% Figure 11:  [4 photos at various phases of CJPL construction .]

\begin{figure}[ht]
  \begin{center}
    \hspace*{1cm}
    \subfloat[]{\includegraphics[width=.39\linewidth]{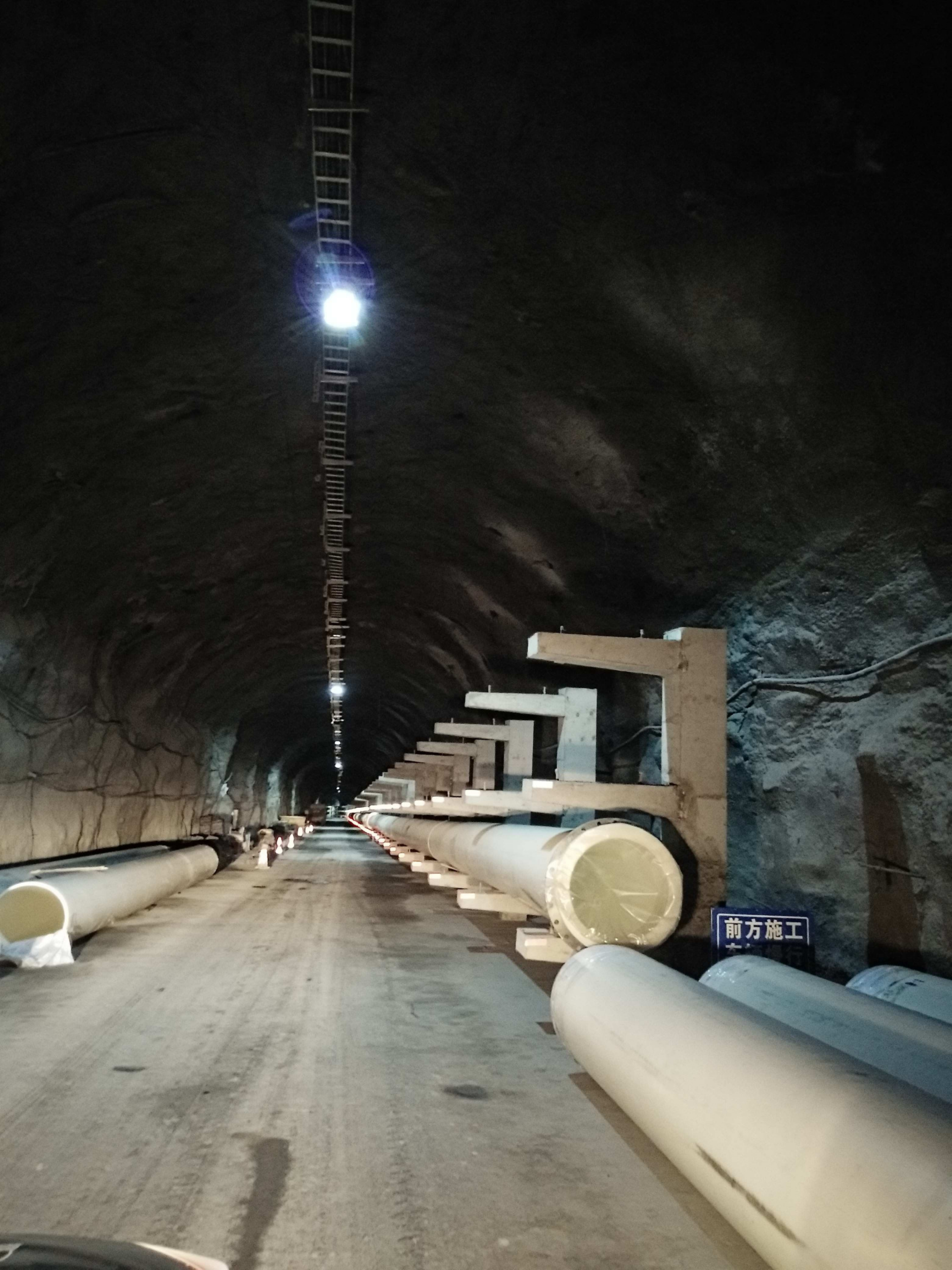}}
    \hspace*{1cm}
    \subfloat[]{\includegraphics[width=.39\linewidth]{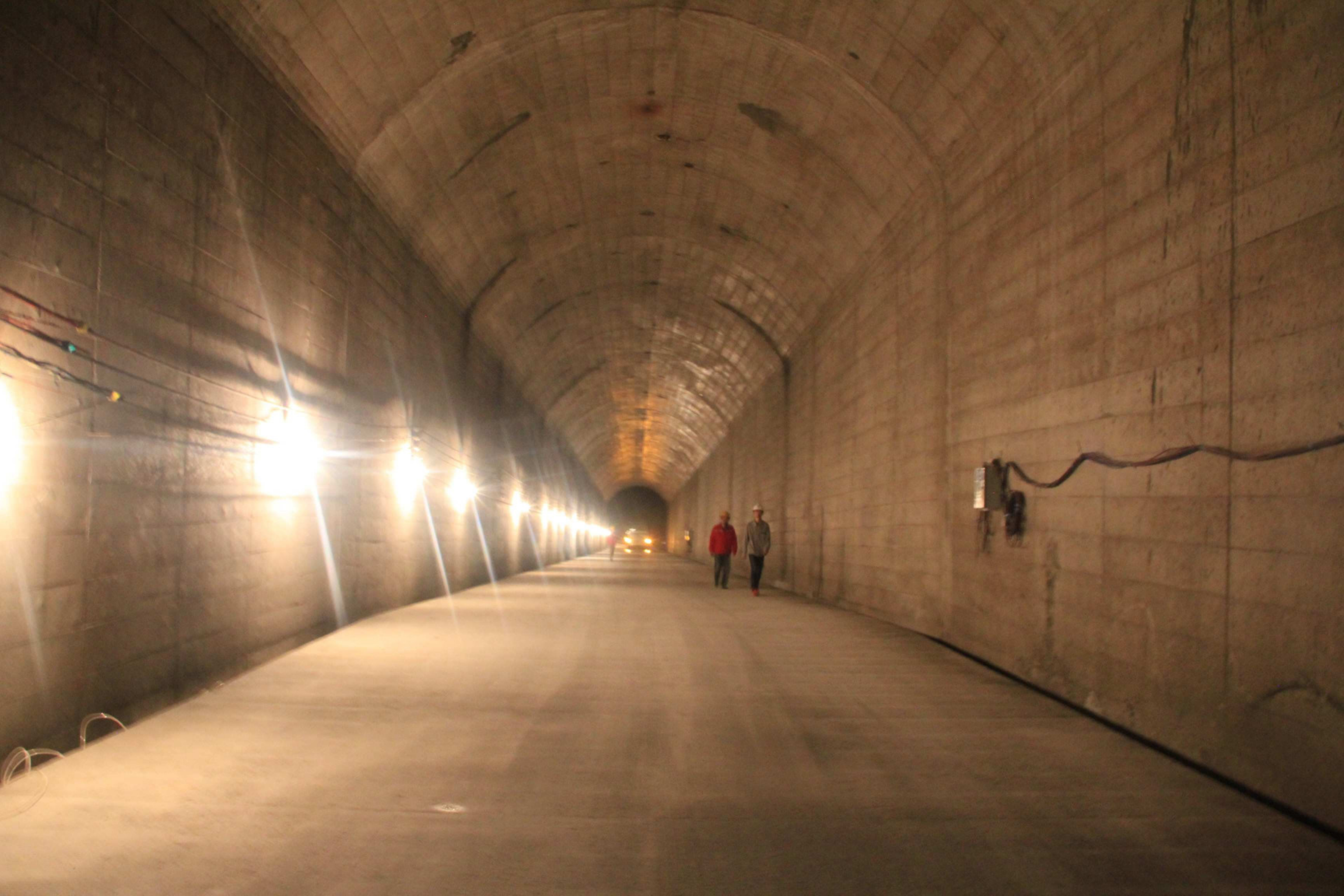}}
    \newline
    \subfloat[]{\includegraphics[width=.39\linewidth]{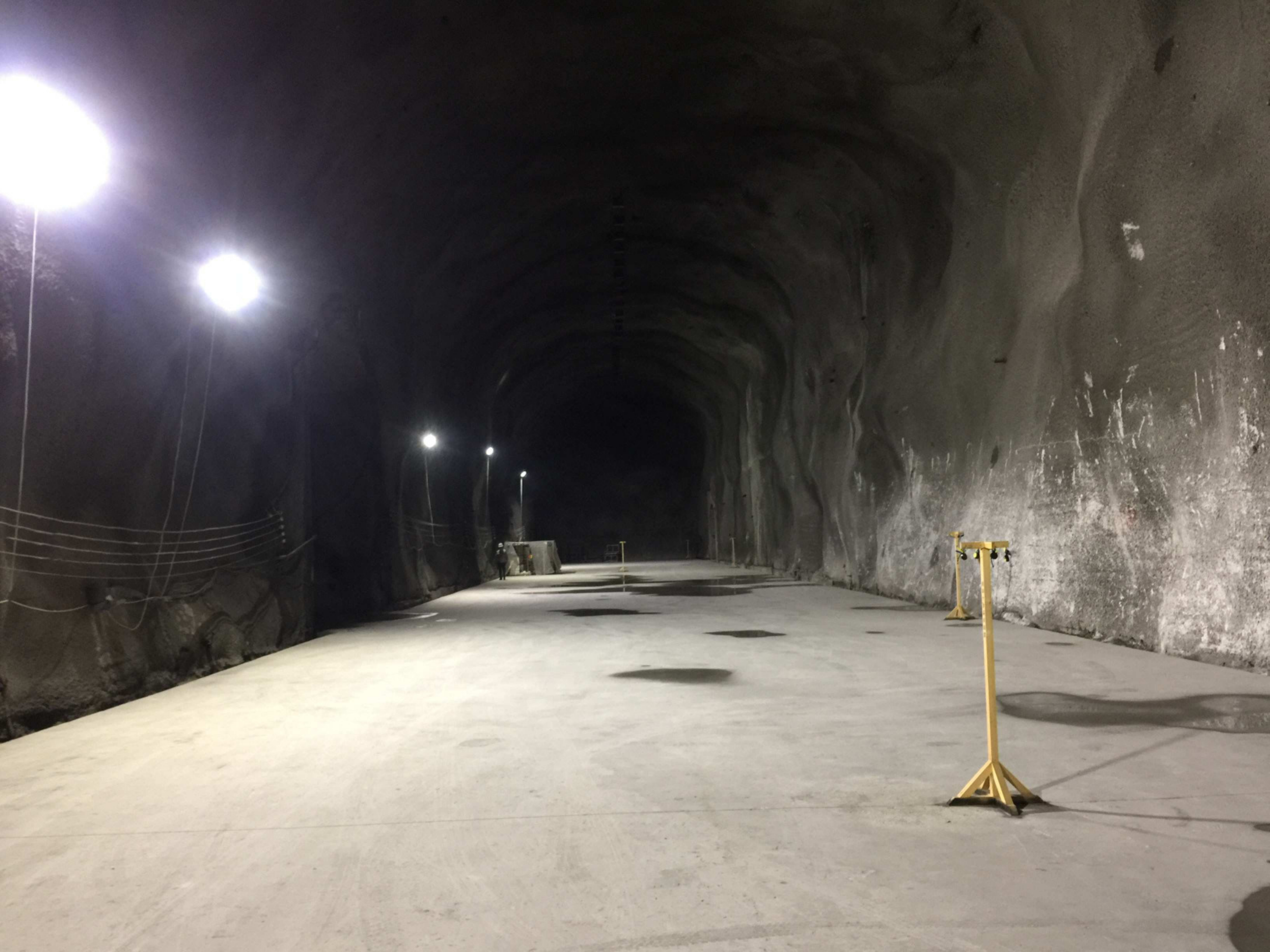}}
    \hspace*{1cm}
    \subfloat[]{\includegraphics[width=.39\linewidth]{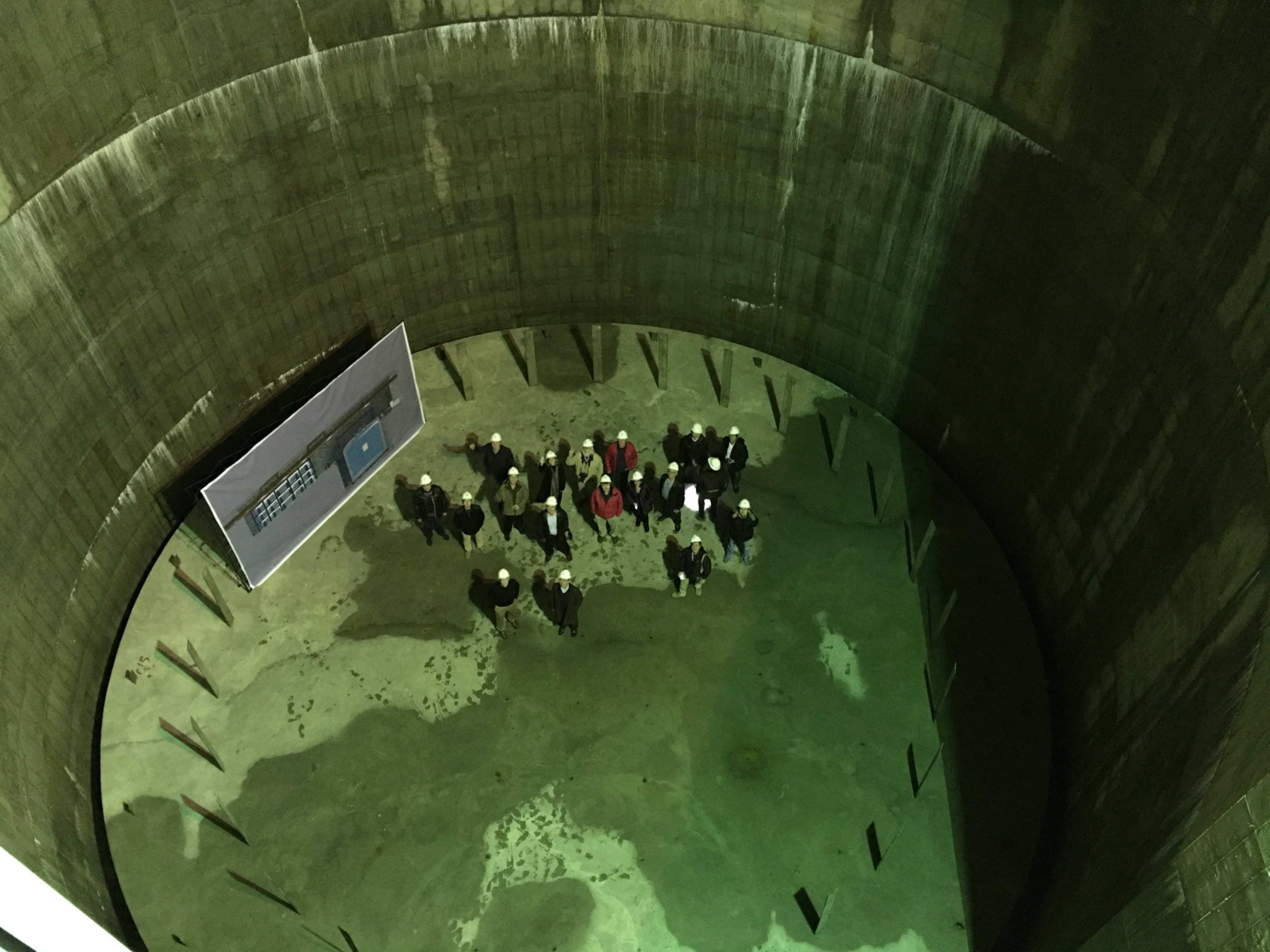}}
  \end{center}
  % =======
  \caption{ Representative photographs during construction of CJPL-II:
    (a) Installation of the ventilation pipes; (b) the connecting road
    tunnel; (c) a view of one of the main hall from its middle section
    towards one end; and (d) the 18~m deep pit of Hall-C.}
  \label{fig::cjpl2_photos}
\end{figure}

%% ===================

Excavation of CJPL-II was performed by the drilling and blasting
technologies. The walls are supported by numerous steel bolts while
the surfaces are finished by custom-selected low-activity
concrete. The potential problems of underground water were alleviated
by numerous embedded 5 cm diameter drain tubes which guide the water
into the ground drainage system. Representative photos during CJPL-II
construction are displayed in Figures~\ref{fig::cjpl2_photos}abcd, the
context of which are described in the caption.

The available electrical power is 10 MW, provided by the Jinping Power
Plant. Ventilation is crucial for human well-being, equipment
performance and radon control. Fresh air will be brought in from the
western entrance of the Jinping Tunnel by three 800~mm diameter PVC
(PolyVinyl Chloride) tubes at specified and maximal flow rates of
${\rm 24000 ~ m^3/h}$ and ${\rm 45000 ~ m^3/h}$, respectively.  The
peak water supply capacity is at ${\rm 1006 ~ m^3/h}$.

The potential science programs at CJPL-II are in various stages of
development. Both CDEX and PandaX have plans to expand and pursue
studies of dark matter and \znu2b{} with detector mass targeted
towards the scale of 1~ton germanium and 10~ton xenon, respectively.
The proposed Jinping Neutrino Experiment~\cite{CJNE:2017ar} will study
solar and geo-neutrinos~\cite{CJNE:2017pr} with a 4~kiloton liquid
scintillator detector, A 1-ton prototype~\cite{CJNE:2017nima} is being
constructed.

%% Paragraph on JUNA  Section 6

The approved JUNA (Jinping Underground laboratory for Nuclear
Astrophysics) program~\cite{JUNA:2016scp1} will benefit from the
ultra-low background conditions of CJPL to apply accelerator-based
low-energy nuclear physics techniques to directly study for the first
time a number of crucial astrophysical nuclear interactions at
energies relevant to the evolution of hydrostatic stars. The
experimental setup includes a high current accelerator based on an ECR
(Electron Cyclotron Resonance) source~\cite{JUNA:2016nima}, advanced
detectors, and low-background shielding systems and will occupy half
of Hall~A of CJPL-II.  The schematic layout is depicted in
Figure~\ref{fig::cjpl2-projects}c. The goals of the first phase of
JUNA are the direct measurements of $^{25}$Mg(p,$\gamma$)$^{26}$Al,
$^{19}$F(p,$\alpha$)$^{16}$O~\cite{JUNA:2016scp2},
$^{13}$C($\alpha$,n)$^{16}$O and $^{12}$C($\alpha$,$\gamma$)$^{16}$O
cross-sections. The experiment will be installed and commissioned on
site in 2018 and the initial series of scientific results are expected
in 2019.

As the deepest underground laboratory in the world, CJPL also offers
an ideal location to study geo-science. Plans have been drawn to build
an international open research platform at CJPL-II to explore the
interaction of deep-rock mechanics with seismology dynamics, to pursue
research on in-situ rock-mass testing methods and pressured coring
techniques, and on advanced theories on rock
mechanics~\cite{GEO:2017aes}. By unveiling characteristics of
irregular constitutive behaviors of deep rock, these studies would
contribute to safe and efficient excavation of mineral resources and
to effective disaster prevention.

Rock excavation and civil engineering of CJPL-II finished in May 2016.
When the construction is fully completed, the facility will become
available to domestic and international researchers.  A dedicated task
force will be responsible to attend the everyday operation and
maintenance of the laboratory.  An international advisory committee
had been setup since 2014 to provide recommendations on the
management, operation and potential science programs.

%% ====================

\section{PROSPECTS AND OUTLOOK}

CJPL, a new underground facility in Sichuan, China, was constructed
within a relatively short period of 8 years. The location has the
deepest rock overburden for cosmic-ray suppression in the world. By
the time CJPL-II is completed in 2017, it will also have the largest
volume of available laboratory space. The facility's horizontal
drive-in access allows efficient deployment of large teams and
equipment. The existing surface support facilities benefit from the
infrastructure created during the earlier construction of hydro-power
projects in the area. The facility will contribute significantly to
the world's capability to conduct low-count-rate experiments.

The two first-generation research programs at CJPL-I, CDEX and PandaX,
focus on searches for dark matter. Competitive results at the world
stage have been achieved. These projects will continue to evolve in
the forthcoming CJPL-II, extending and expanding the scope of their
physics programs to include other areas of study, such as \znu2b{}. A
new program on low-count-rate nuclear astrophysics measurements, JUNA,
is in preparation.

There remains plenty of space for prospective future users at CJPL-II.
The science programs are still in the formative stage. The facility
has attracted a great deal of user interest, both nationally and
internationally. The research community looks forward to the evolution
of the ideas and projects under way at CJPL.

%%  ======= This is not necessary in an arXiv article ==========
%% \section*{DISCLOSURE STATEMENT}

%% The authors are not aware of any affiliations, memberships, funding,
%% or financial holdings that might be perceived as affecting the
%% objectivity of this review.

%% ============================

\section*{ACKNOWLEDGMENTS}

The authors thank all of the scientific colleagues and technical staff
who contributed to the realization of the facility and science
programs reported in this review. Construction and operation of CJPL
are financed by the National Natural Science Foundation of China
(11355001), and basic facility funding is provided by the Ministry of
Education of China. Funding support for the CDEX program is provided
by the National Natural Science Foundation of China (11275107,
11355001, 11475117, and 11475099) and the National Basic Research
Program of China (973 Program) (2010CB833006). The PandaX program is
supported by grants from the National Science Foundation of China
(11435008, 11455001, 11505112, and 11525522). We greatly appreciate
the contributions of Xun Chen, Xia Li, Shin-Ted Lin, Shu-Kui Liu, and
Li-Tao Yang in the preparation of this review.
This article is posted with permission from the
{\it Annual Review of Nuclear and Particle Science},
Volume 67 \copyright ~ 2017
by Annual Reviews, http://www.annualreviews.org/.

\end{document}